# Reusing empirical knowledge during cloud computing adoption

Mahdi Fahmideh, Ghassan Beydoun

Faculty of Engineering and Information Technology, University of Technology Sydney
Mahdi.Fahmideh@uts.edu.au, Ghassan.Beydoun@uts.edu.au (corresponding author)

**Abstract.** Moving legacy software systems to cloud platforms is an ever popular option. But, such an endeavour may not be hazard-free and demands a proper understanding of requirements and risks involved prior to taking any actions. The time is indeed ripe to undertake a realistic view of what migrating systems to the cloud may offer, an understanding of exceptional situations causing system quality goal failure, and insights on countermeasures. The cloud migration body of knowledge, although is useful, is dispersed over the current literature. It is hard for busy practitioners to digest, synthesize, and harness this body of knowledge into practice in a scenario of integrating legacy systems with cloud services. We address this issue by creating an innovative synergy between the approaches *evidence-based software engineering* and *goal-oriented modelling*. We develop an evidential repository of commonly occurred obstacles and platform agnostic resolution tactics related to making systems cloud-enabled. The repository is further utilized during the systematic goal-obstacle elaboration of given cloud migration scenarios. The applicability of the proposed framework is also demonstrated.

## 1. Introduction

Cloud computing is a fundamental shift in delivering IT services to software systems. A perennial concern of IT managers embarking on migrating critical legacy systems to cloud platforms is to ensure attainability of their goals (Khajeh-Hosseini, Sommerville, Bogaerts et al. 2011). Despite pervasiveness and hype over cloud computing, some organisations are still reluctant to undertake migration projects. Whether the migration is of legacy systems to the cloud or changing an existing cloud platform, perceived uncertainties often hinder undertaking such projects. Uncertainties originate from various factors, e.g. data security, failure accounts of other organisations, vendor lock-in in absence of standards, cultural shift, unclear jurisdiction for online activities over distributed cloud data centres, service outage, and many others (Chow, Golle, Jakobsson et al. 2009; Pepitone 2011; Linthicum 2012; Tsidulko 2016). For example, the reliability of cloud services is sometimes questioned because of the outage of Google GMail service or Microsoft's Danger division's causing loss of some customers' data. Failure to adequately identify and mitigate such risks beforehand may become costly to rectify if they are detected at later stages when systems are in operating in the cloud. Ideally, such issues should be accounted for requirement analysis time when system goals are being identified. This would allow more flexibility to negotiate multiple trade-offs and can lead to a cheaper overall outcome in a satisfactory way.

Since the emergence of cloud computing technology in 2007, there has been an ever increasing number of versatile accounts, published by both academia and industrial ends, on effective adoption of cloud services to augment operation and maintenance of legacy systems in different organisational and project settings. Such documented accounts provide a test bed that can be reused for informed decision making in moving systems to or across cloud platforms. Nevertheless, given the widespread of the literature produced, a systematic support that capitalizes this body of knowledge to make it more explicit, reusable, and accessible is non-extant yet.

Repeated calls by (Giovanoli 2012; Zimmermann, Wegmann, Koziolek et al. 2015) have remained largely unheeded for capturing and reusing cloud migration knowledge to improve decision making which subsequently has an impact on various system quality goals. This article alleviates this gap via deploying a combination of *evidence-based software engineering* (EBSE) (Dyba, Kitchenham and Jorgensen 2005) and *goal-oriented modelling* (Yu 1997). We introduce a knowledge-based decision support framework that systematises reusing the existing body of cloud migration knowledge. The framework comprises an evidential repository of commonly occurring cloud migration goals, obstacles hindering satisfying cloud migration goals, and corresponding countermeasures to handle these obstacles. The repository has been identified through an extensive review of published studies and experience reports in the literature. The repository information is further utilized during reasoning about requirements of cloud migration scenarios. We believe the proposed framework helps a system architect in better handling of potential risks before they are propagated in later stages of cloud migration and thus improving the reliability of decision outcomes. We illustrate the applicability of the framework in two scenarios.

The rest of this article is organised as follows. Section 2 presents a motivating scenario for this study. Section 3 presents the research methodology conducted to develop and validate the proposed framework. Section 4 delineates the development of the framework components. Section 5 illustrates the application of the framework in two scenarios of moving legacy systems to cloud platforms. In the view of a set of analysis criteria, Section 6 reviews related work. Section 7 provides discussion on the benefit of using the framework, following with validity threats in Section 8. Finally, section 9 includes the research summary, conclusion, and future research directions.

## 2. Motivating scenario

The uncertainty surrounding cloud enablement of legacy systems may raise some challenges. Our research is inspired by a real-world cloud migration scenario in oil and gas industry sector discussed in (Khajeh-Hosseini, Greenwood and Sommerville 2010) through which an IT solution organisation moves a legacy system from an in-house data centre to Amazon EC2. The system allows users, who own an oil rig located in the North Sea oilfields, to manage, monitor, and acquire minutely data from an off-shore oil rig operations. The system comprises a database layer that logs and archives data coming from offshore in a database and tape for taking daily database backups. The business logic layer provides functionalities for data reporting and monitoring. The end users access the system through using a remote desktop client over the internet. The real-time data that are coming from onshore are provided for users via communication links provided by the IT solution organization. The organisation has responsibility for maintaining and upgrading the system.

Top level management of the IT solution organization intends to augment the scale of servicing and competitiveness via expanding its system services to users who own Middle-east oil rigs. Nevertheless, the organization cannot afford procurement and maintenance of new infrastructure to support timely processing of upcoming massive scale data from multiple oil rigs during the workload. Cloud services attract the top level management as they are said to provide powerful infrastructures along with a wide-range of services. A system architect is appointed to design an overall architectural solution to deploy the system in Amazon EC2 Web services as a co-location. However, she is unsettled with many intriguing questions being asked by the top level management, for example:

(i) By moving these systems to the cloud, will higher system performance be attainable in all situations?
(ii) What risks are likely to obstruct reducing infrastructure cost and system security in the cloud? and
(iii) How such risks can be negated in advance?

If the system reengineering to cloud is to achieve its potential, this sort of questions should be clearly answered. The system architect might have basic knowledge of promised benefits and issues around migrating systems to the cloud. However, she may face difficulty in making informed answers to the abovementioned questions due to uncertainty and little objective evidence to confirm suitability and inherent risks of such transition. For instance, the choice of replacing the current system relational database to a new non-SQL cloud database solution may have an uncertain impact on the query processing time and thus system throughput. She may seek and select various information resources such as documents, weblogs, domain expert advice, or personal experience. Due to voluminous such sources in the cloud computing field, in particular the continuous growing publication rate since 2008 (Yang and Tate 2012), it becomes more cumbersome for her to grasp, synthesise, and reuse extant material for the given scenario since they may not be easy to find among the mix of other papers. Furthermore, she may rarely review or even have access to them. Solely, if they are collected, the system architect may not be able how to analyse these contents and envisage implications to organisational strategic goals.

EBSE approach is known to be one of most successful solution for an informed decision on a new technology adoption. In the spirit of EBSE, best pieces of evidences from scientific publications are capitalized by collecting, generalizing, documenting, and storing in an evidential repository which can be later reused for a decision making situation (Dyba, Kitchenham and Jorgensen 2005). In this research, we provide an evidential repository assorting the most evidential goals, obstacles, and countermeasures on how to negate obstacles.

Our objective is not only to develop an evidence-based repository, but also to utilize the repository and incorporates its information in the suitability assessment of given cloud enablement scenarios. Reusing the repository requires a systematic support that models and processes the information in the repository in association with a variety of parameters e.g. goals, risks, effort, size, or calendar time. We settled on the goal-oriented modeling approach for exploring the repository to make informed answers to the abovementioned questions. Goal-obstacle analysis explicitly relates high-level cloud migration goals with potential obstacles and relevant countermeasures addressing these obstacles. Little or no research has focused on how the early stage suitability analysis of cloud enablement can be complemented in the presence of evidential data available in the literature.

## 3. Research methodology

This research pursuit is to craft an IT artefact. The research paradigm that suits this inquiry is design science research (DSR) (Henver, March, Park et al. 2004) through which a viable artefact addressing a relevant solution to an unsolved problem is developed and validated. We conducted three phases of a typical DSR, but tailored for the purpose of this research, as delineated in the following:

*Phase 1- Problem identification* has been already described in the sections 1, i.e. the lack of a systematic knowledge reuse to improve the reliability of goal-obstacle analysis results and decision outcomes. Our research objective is set up as “developing of a systematic framework reusing empirical evidence for goal-obstacle analysis at the early stage of migrating systems to the cloud”.

*Phase 2- Design and development* of the framework that constitutes the development of two core components as follows:

- (i) an empirical knowledge repository of recurring goals, obstacles, and countermeasures in cloud enablement of legacy systems,
- (ii) a procedure including steps to identify cloud migration goals, potential obstacles, assessing their risk, i.e. likelihood and severity, and tackling them by generating new goals.

In the design science research, different approaches and kernel theories from inside or outside of the software engineering discipline informing an artefact creation can be brought to bear (Gregor and Jones 2007). As mentioned earlier, for the development of the first component, i.e. the knowledge repository, we employed EBSE approach (Dyba, Kitchenham and Jorgensen 2005). A common technique to run EBSE is Systematic Literature Review (SLR) (Kitchenham, Pearl Brereton, Budgen et al. 2009) where findings from different empirical studies are gathered and summarized regarding inclusion criteria and indicators to draw plausible conclusions (Kitchenham, Pearl Brereton, Budgen et al. 2009). In this research, the framework’s repository has been developed out of an SLR of published works empirical studies in the cloud migration literature.

For the second component, i.e. the procedure, we employed a generic goal-oriented modelling framework called KAOS (Keep All Objects Satisfied). KAOS provides support for elaborating, structuring and analysing software requirements, including both functional and non-functional ones (Van Lamsweerde 2009). It also supports different levels of expression and reasoning that vary from semi-formal to formal analysis goal models depending on the reasoning precision sought (Dardenne, Van Lamsweerde and Fickas 1993; Van Lamsweerde and Letier 2004). In KAOS, goals are iteratively refined through top-down (by asking *how* questions to refine goals into sub-goals) as well as a bottom-up way (by asking *why* questions to identify parent goals). The refinement proceeds until all goals reach clear and assignable responsibilities to agents who realize the goals. We used this modelling framework in conducting the cloud enablement goal-obstacle analysis.

Generic KAOS's concepts such as goal, obstacle, and resolution tactic do not provide precise definitions that can be refined into testable and operational cloud migration requirements. For example, KAOS's concept obstacle refers to "an exceptional condition that prevents a goal from being satisfied" (van Lamsweerde and Letier 2000). In our view, a preliminary use of KAOS that specifies and refines high-level goals would not be sufficient. There is no operational definition for a refinable and testable cloud-related goal-obstacle analysis. We enriched KAOS's generic concepts with cloud-specific knowledge provided by the evidential repository. For instance, the generic notions of obstacle and resolution tactics in KAOS have been augmented with 67 and 45 cloud-specific obstacles and resolution tactics, respectively. This will be later detailed in Section 4.1.

*Phase 3 - Validation* appraised the efficacy of the framework resulting from phase 2 through (i) a Web-based survey and (ii) two case studies of moving an open-source Web-based legacy system, providing real-time stock quotes, to Pivotal Cloud Foundry and a digital document processing legacy system to Microsoft Azure cloud platform.

DSR is an iterative develop-and-validate process in the sense that the developed artefact is situated in a problem space and is iteratively refined to fulfil quality and utility metrics (Henver, March, Park et al. 2004; Peffers, Tuunanen, Rothenberger et al. 2008). In the context of this research, we conducted two consecutive cycles.

The first cycle took place between February 2014 and September 2016. It resulted in the initial version of the framework including its repository and procedure. The collection of obstacles and resolution tactics were validated using experts in the SLR, goal modelling, and cloud computing areas. In addition, the resolution tactics were validated for completeness through a comparison against existing migration methods to verify if they are sufficiently complete (Fahmideh 2016b) and through an expert review using a public Web-based survey questionnaire (Fahmideh, Daneshgar, Beydoun et al. 2017). The survey examined if the resolution tactics are perceived as important and relevant for incorporating into the process of legacy system reengineering to cloud platforms. We used purposeful sampling (Patton 1990) to identify eligible experts to participate from their public profiles on social media such as Linkedin, Twitter, and academic research groups. Domain experts were contacted by e-mail to confirm their expertise. Once willingness and expertise were confirmed, an invitation along with the link to the survey was issued. Experts were asked to rate the importance of each resolution tactic on the basis of a seven scales (1–7) where 1 represents 'completely irrelevant', 2 indicates 'unimportant', 3 for 'somewhat unimportant', 4 for 'neither important nor unimportant', 5 for 'somewhat important', 6 for 'important', and 7 for 'extremely important'.

In this voluntary survey, we invited 515 experts but 144 experts answered the survey. After removing incomplete responses, 104 answer-sheets were used for data analysis. The respondents were from 32 countries with an average of cloud migration experience of 3.8 years. The statistical analysis of responses revealed that the majority of the tactics in the repository were perceived sound and important for incorporation into the cloud migration projects. More detailed can be found in (Fahmideh, Daneshgar, Beydoun et al. 2017).

In the second DSR cycle, between September 2016 and December 2016, we specialized and adapted two real-word case studies available in the literature. The first scenario, named *SpringTrader* (Gordon 2015), described moving an open-source stock screener Web-based system to Pivotal Cloud Foundry platform. In the second scenario, named *InformIT* (Rabetski 2012; Rabetski and Schneider 2013), a digital document processing system was migrated to Microsoft Azure platforms. For both case studies, we used secondary documents obtained from a variety of sources, mainly SpringTrader's Weblog and 43-page project documentation of *InformIT* project, to enrich our understanding of the enacted migration process model including project sequence, the architecture of the legacy system and cloud solution, and user histories. We traced the projects' documents if and what goals each scenario defined, issues that were occurred and countermeasures that were applied. The detail description of the scenarios is presented in Section 5.

In this article, we only present the development of the repository and the procedure components in the first cycle and validating the applicability of the framework in the second cycle.

## 4. Development of the framework

As shown in Figure 1, the framework comprises (i) a repository holding collections of obstacles and resolution tactics, and (ii) a goal-obstacle analysis procedure relying on the repository. In Section 4.1, the result of the literature review to develop the framework repository is presented. This includes establishing a literature review protocol, conducting the review, and identifying, synthesising, and organising the collections. Section 4.2, presents the proposed goal-obstacle analysis procedure.

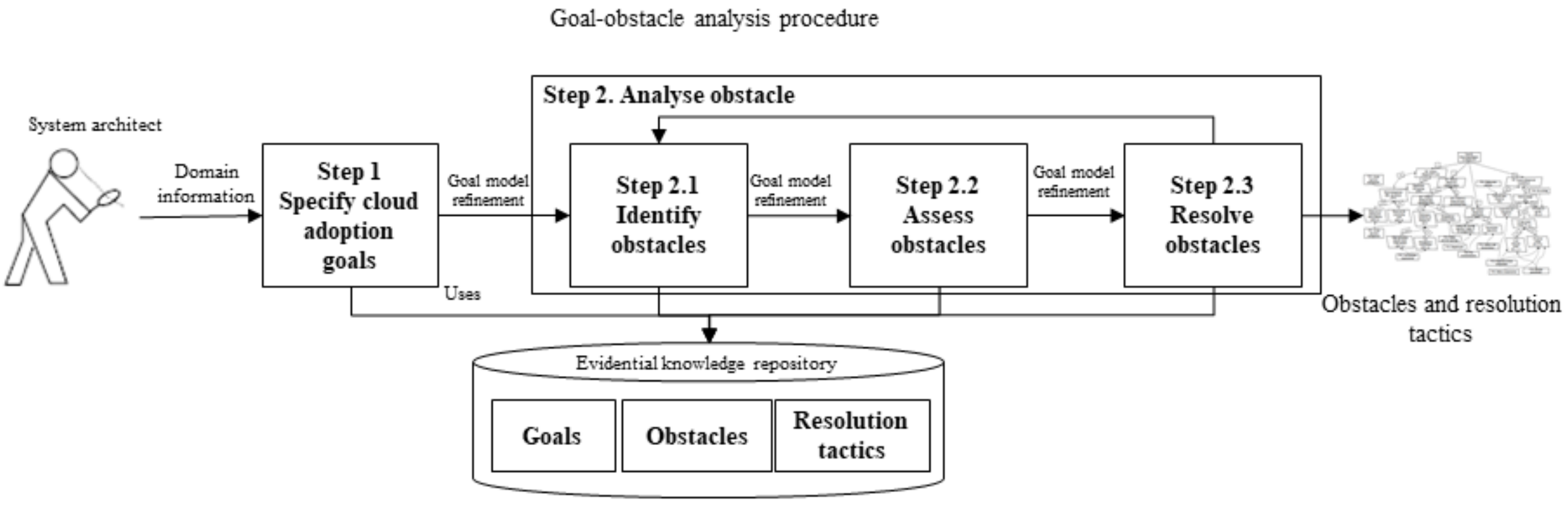

Figure 1. Structure of the proposed framework

### 4.1 Developing the Repository

Three research tasks were defined to develop the repository's collections:

(i) Derivation of common quality goals expecting to be satisfied by migrating systems to the cloud
(ii) Derivation of recurring obstacles against achieving quality goals, and
(iii) Derivation of resolution tactics in handling these obstacles

Figure 2 depicts the SLR was undertaken to conduct tasks (ii) and (iii). We did not reckon a need to follow an SLR for task (i) as we used a fixed set of system quality goals commonly agreed in software engineering and cloud computing literature. The objective of the SLR was to answer the following inquiries: (i) *what obstacles may occur against system quality goals when moving systems to or they are in operation in the cloud* and (ii) *what resolution tactics are available to address the obstacles?* The same SLR was conducted for tasks (ii) and (iii).

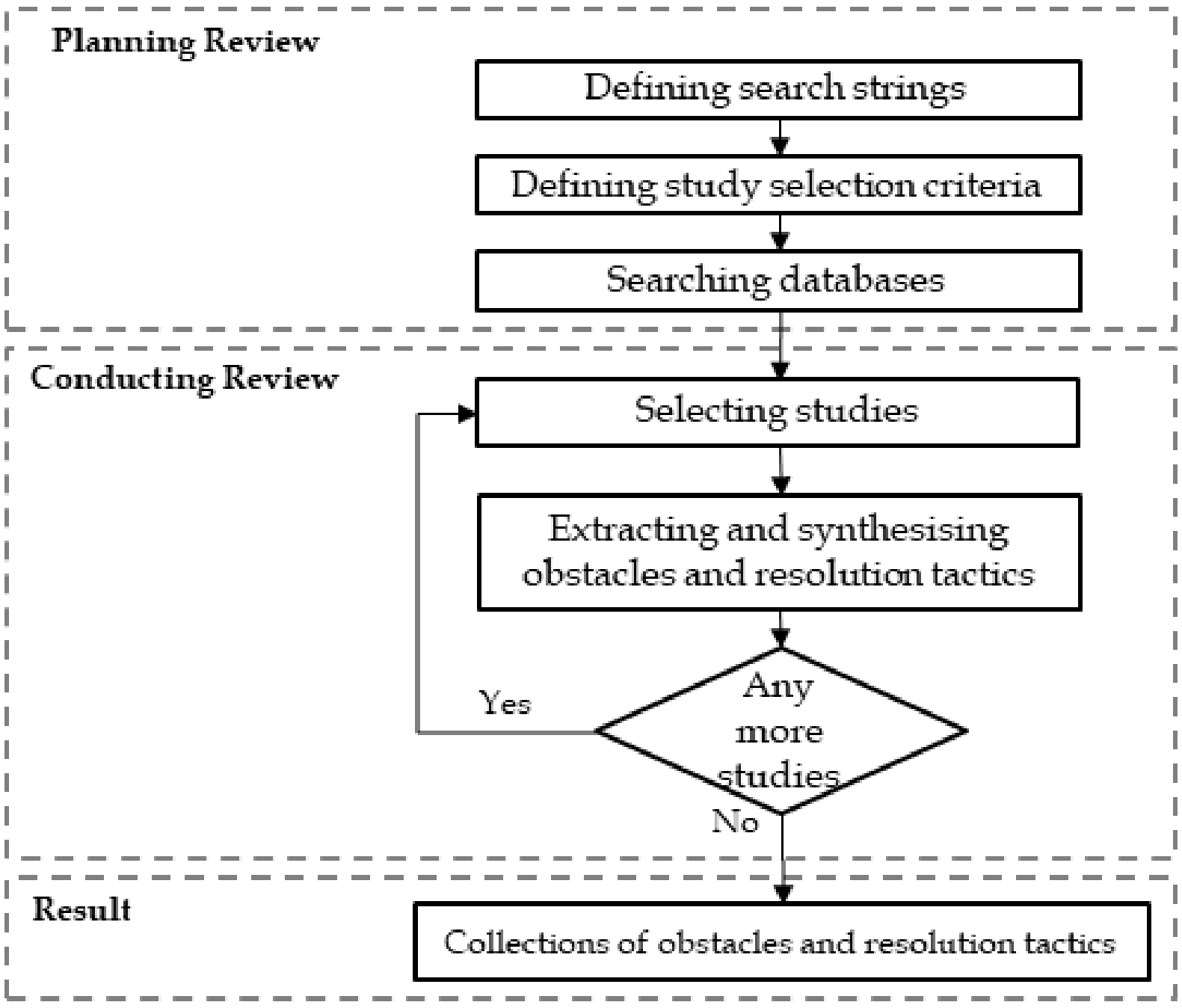

Figure 2. SLR conducted for developing the framework repository – duration between February 2014 and September 2016

### *Planning Review*

The objective of this phase was to tackle any researcher bias (Kitchenham, Pearl Brereton, Budgen et al. 2009) through defining search strings, study selection criteria, and searching databases.

**Defining search strings.** The search strings were defined based on the guidelines recommended in (Dieste and Padua 2007). These included: (i) defining main terms by breaking down the research questions, (ii) identifying alternative synonyms for main terms, (iii) checking the search strings in any relevant papers that retrieved, (iv) incorporating alternative synonyms using the logical *OR* and *AND* operators to link the main terms. The terms *cloud computing*, *legacy*, *reengineering*, *migration*, and *IaaS*, *PaaS*, and *SaaS* were set as the main keywords from which different search strings defined and combined using the operators OR and AND. Table 1 shows some examples of generated search strings.

Table 1. list of related search strings (SS)

| |
|---|
| SS1: "Migration" OR "Cloud adoption" OR "Cloud migration" OR "Migration to cloud" OR "Legacy to cloud migration" OR "Legacy migration to cloud" AND [SS2 OR SS3 OR SS4 OR SS5 OR SS6] |
| SS2: "IaaS risks" OR "IaaS challenges" OR "IaaS challenges" OR "IaaS adoption" OR "IaaS benefits" |
| SS3: "PaaS risks" OR "PaaS challenges" OR "PaaS issues" OR "PaaS adoption" OR "PaaS benefits" |
| SS4: "SaaS risks" OR "SaaS challenges" OR "SaaS issues" OR "SaaS adoption" OR "SaaS benefits" |
| SS5: "Monolith application" OR "Legacy code" OR "Legacy system" OR "Existing system" OR "Legacy component" OR "Legacy software" OR "Legacy application" "On-premise application" OR "Monolithic system" OR "Existing software" OR "Pre-existing software" OR "Legacy information system" OR "Legacy program" OR "Pre-existing assets" OR "Legacy architecture" OR "Legacy asset" |
| SS6: "Reengineering" OR "Legacy system reengineering" OR "System reengineering" |

**Defining study selection criteria.** From the identified studies those selected that (i) were related to migrating or developing systems to/for cloud platforms with a proper description of the context and clear objectives (ii) described situations, i.e. obstacles, that may cause goal failure and if any resolution tactics, (iii) provided a proper validations through case study, example, interview, etc., (iv) published from 2007 onwards in software engineering and information systems journals/conference proceedings, and (v) described in English language.

**Searching databases.** The following databases were searched: IEEE Explore, ACM Digital Library, SpringerLink, ScienceDirect, Wiley InterScience, ISI Web of Knowledge, and Google Scholar. We also did not overlook internet blogs and trade journal articles provided empirical accounts for the specific platform such as Amazon.

### *Conducting Review*

**Selecting studies.** The databases listed in the previous step were searched using the search strings. The whole content of each identified study was screened regarding the inclusion criteria. It is important to mention that conducting the review was not a linear and mechanical process; rather it was a hermeneutic, iterative, and informed by careful reading each study and understanding its context. Forward and backward searches were performed so that studies cited in the references and related work sections of the study were fed into this step to find new studies. The review phase, strictly speaking is open-ended, resulted in identifying 112 studies as shown in Appendix A.

**Extracting and synthesising obstacles and resolution tactics.** Each study' segment that stated any obstacles or resolution tactics were extracted along with the reference to the study. Some leading questions that were used during the development of the collections were as follows: (i) does the study report any technical or social obstacles that may cause cloud adoption goals fail? If so, what is the obstacle? (ii) how can the obstacle influence the successful adoption of cloud services? and (iii) Are there any resolution tactics suggested by the study to overcome the obstacles? The collections obtained through this step presented in Appendix B. A synopsis is provided herein what follows:

***(i) Goal collection*** includes ten ready-made common software system quality goals that cloud services can positively contribute to the efficiency of legacy systems. This includes *Availability, Scalability, Security, Performance, Customizability, Interoperability, Portability, Testability, Consistency*, and *Reduced IT cost*. These goals facilitate initialization and refinement of goal models as described in Section 4.2.

***(ii) Obstacle collection*** has information about 67 common probable, technical or social, situations causing quality goal failure and thus hampers systems benefit from cloud services.

***(iii) Resolution tactic collection*** contains 45 platform-agnostic solutions applicable for handling obstacles. Resolution tactics are a result of applying abstraction and synthesisation to existing ad-hoc implementation techniques to utilize cloud service available in the literature. Our framework uses them during the goal-obstacle analysis to explore alternative ways to resolve obstacles. Note that this research tended to keep resolution tactics at the abstract level. Thus their operationalization details are left to developers or manager as to existing supportive techniques or tools available in the cloud computing marketplace.

Based on the common service delivery models IaaS, SaaS, and PaaS, Fahmideh et. al. defines a few variants through which legacy systems can utilize cloud services (Fahmideh, Daneshgar, Low et al. 2016). These are defined as follows. In *Type I*, the business logic layer of a system, which offers discrete and reusable functionality, is deployed in cloud infrastructure through IaaS model such as Amazon EC2 but the data layer is kept in an on-premises network. In *Type II*, system components are replaced with fully tested cloud services using SaaS model. In *Type III*, the system database is deployed in a cloud data store provider such as Amazon Simple Storage Service (S3), Amazon Elastic Block Store, Dropbox, or Zip Cloud whilst business logic components are maintained on an on-premises network. In *Type IV*, the database of a legacy is modified and converted to a cloud database solution such as Amazon SimpleDB, Google App Engine data store, or Google Cloud SQL. Finally, in *Type V* the whole system stack is encapsulated in virtual machines and ran on servers.

Adopting each abovementioned migration types may face some obstacles. For example, it is quite common for incompatibility issues between legacy system data type and a chosen cloud database solution to arise in the case of adopting migration types I, II, IV, and V. To indicate such situations, the collection of obstacles in Appendix B shows if an obstacle is related to a migration type via symbols √. During step 2.1 of the goal-obstacle analysis procedure, this information is used to identify obstacles.

**Result.** Table 2 shows an excerpt of the information stored in the repository. Each of goal, obstacle, and resolution tactic is respectively denoted by an identifier-number G, O, and T. For example, the quality goal for a system is that it should be *interoperable* (G6) across different platforms. Studies [S2], [S3], [S4], [S5], [S35], [S36], and [S37] mention that cloud services are supposed to be *interoperable* (G6) across different platforms and integrable with systems. Nevertheless, there are some potential obstacles obstructing the interoperability goal. These obstacles, for example, as evidenced in [S23], [S24], [S12], [S38], [S25], [S26], [S27], [S39], and [S40] are *Incompatible pluggable cloud services (O19)*, *Incomplete APIs (O20)*, *Incompatible datatypes (O21), Operating system incompatibility (O22)*, and *Machine-image incompatibility (O23)*. In addressing these potential obstacles, two generic tactics *Refactor legacy source code (T2)* and *Develop adaptor/wrapper (T3)* are suggested in [S65], [S66], [S67], [S75], [S76]. Maintaining consistency, the architect may slightly change the original names of these goals, obstacles, and resolution tactics for simplifying modelling.

Table 2. An excerpt of probable obstacles obstructing the quality goal *system interoperability* along with some alternative resolution tactics and reference to the empirical studies

| **Quality goal** | **Definition** | **Source** |
|---|---|---|
| G6 | *Interoperability*. Cloud services can be illimitably incorporated to and integrated with the systems. | Genera literature on cloud computing (e.g. [S2], [S3], [S4],[S5], [S35], [S36], [S37] |
| **Obstacle** | **Definition** | **Source** |
| O19 | *Incompatible pluggable cloud services*. At runtime, system might be plugged to a cloud service which is incompatible with the other cloud services. | [S23] |
| O20 | *Incomplete APIs*. Cloud service provider lacks providing a rich set of APIs. | [S24] |
| O21 | *Incompatible data types*. Data types used in legacy and cloud service are incompatible. | [S12], [S38] |
| O22 | *Operating system incompatibility*. System components are distributed and moved among cloud servers with different operating systems which might be incompatible for managing, representing, and formatting virtual machines. | [S25], [S26], [S27] |
| O23 | *Machine-image incompatibility*. Virtual machines are moving between different cloud platforms but each platform has different underlying implementation for virtual machines. | [S39], [S40] |
| **Resolution tactic** | **Definition** | **Source** |
| T5 | *Refactor legacy source code*. Modify the system source code for being compatible and be able to interact with selected cloud platform programming language and APIs. | [S65], [S66], [S67] |
| T3 | *Develop adaptor/wrapper*. Add adaptors for resolving mismatches, occurring at runtime execution, between legacy system components and cloud services. | [S75], [S76] |

## 4.2 Establishing goal-obstacle analysis procedure

As mentioned earlier, this research uses KAOS modelling concepts to elicit, model, and reason about goals for migrating systems to cloud platforms. We enriched KAOS generic concepts by introducing 67 obstacles and 45 tactics. Table 3 presents KAOS modelling concepts used for the second component.

The procedure includes the following two steps: (i) *Specify cloud migration goals* to set up and visualize high-level quality goals targeted for moving systems to the cloud, (ii) *Analyse obstacles* comprising sub-steps for identifying obstacles causing goal failure, assessing their risk, and defining resolution tactics to modify existing goals, or generating new ones to prevent, to reduce, or to mitigate the obstacles. The output of the procedure is a consolidated requirement model representing cloud migration goals, potential obstacles to be tackled, and (alternative) resolution tactics. This model can be later incorporated into the system implementation phase.

To illustrate the inner working of the procedure, an example scenario of moving the database of a legacy system to the cloud service Amazon Simple Storage (S3) (AmazonS3), a public, secure, and highly scalable data storage), is described. This is an instance of migration type V.

Table 3. notations used for goal modelling

| Modelling element | Definition | Graphical notation |
|---|---|---|
| Migration type | An option through which a system can benefit from cloud services to improve its working performance (See section 2). | 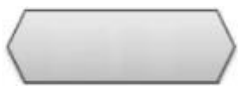 |
| Goal | A quality goal that is expected to be satisfied by adopting cloud services. | 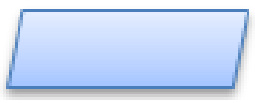 |
| Obstacle | A technical or a none-technical exceptional situation/condition preventing the goal satisfaction. | 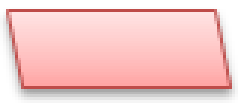 |
| Resolution tactic | A generic solution (.i.e. new goals, assumptions, or by modifying existing goals) to resolve an obstacle. | 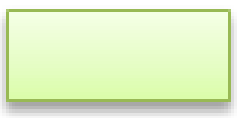 |
| Decomposition | A mechanism to refine a goal/obstacle to a set of fine-grain goals/obstacles. | 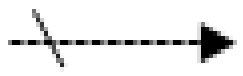 |
| Contribution | Positive contribution of migration type to a quality goal. | 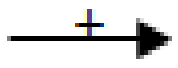 |

### Step 1 Specify cloud migration goals

The framework provides a collection of pre-defined quality goals commonly intended in moving legacy systems to the cloud that the system architect and stakeholders can use to initiate a goal model. In this scenario, three goals are set for moving the system database to S3 platform (Figure 3). This includes *Achieve [Reduced IT cost]*, *Achieve [Improved performance]*, and *Achieve [Improved availability]*. Goals can be decomposed into fine granular ones for more accurate analysis. The goal *Achieve [Improved performance]* is a combination of sub-goals *Achieve [Reduced data uploading time]* and *Achieve [Reduced query processing time]* meaning that the satisfaction of *Achieve [Improved performance]* depends on the satisfaction of both these sub-goals. In Figure 3, the dotted arrows show the fact that goals, obstacles, and resolution tactics are extracted from the repository.

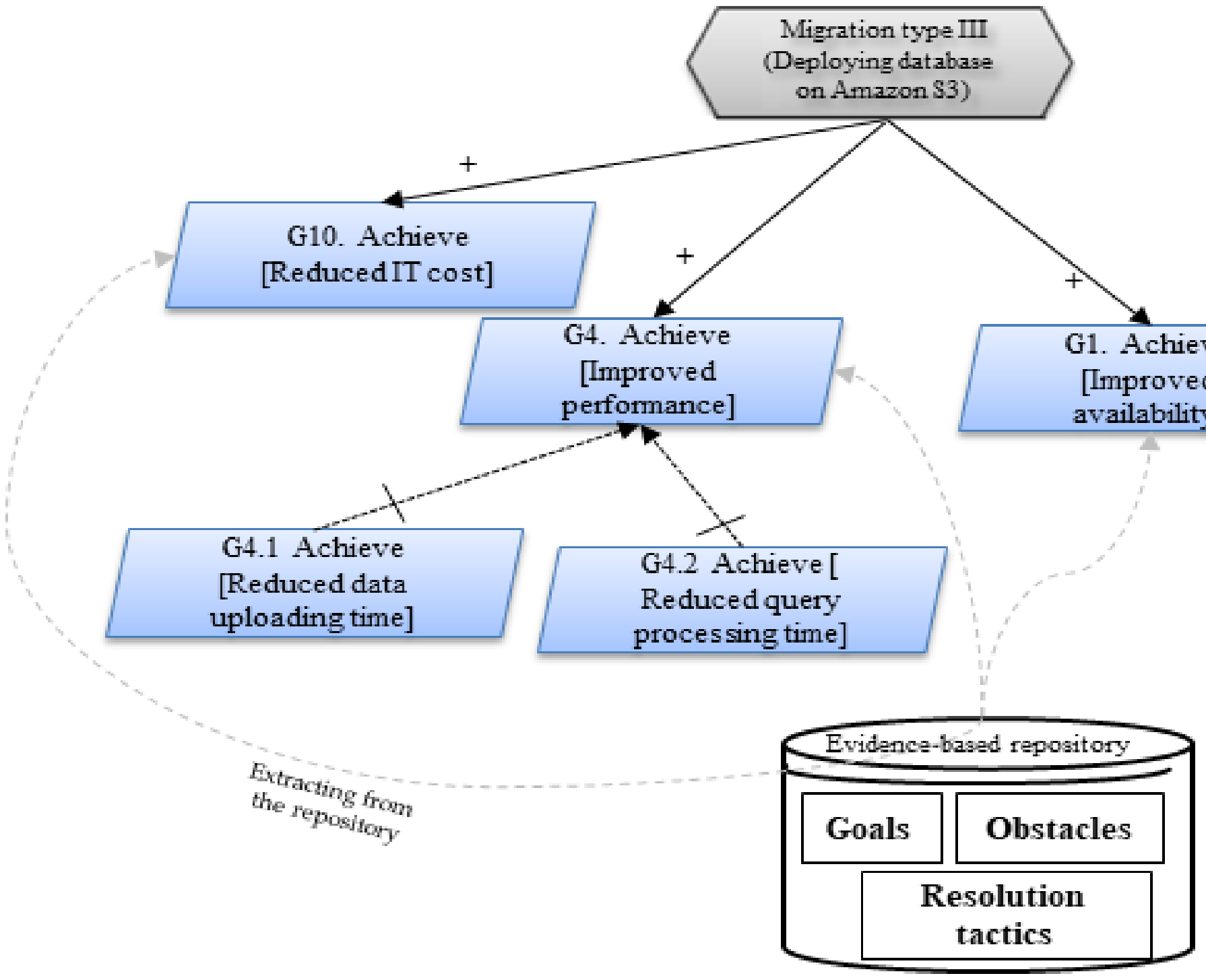

Figure 3. Goals in moving the system database to Amazon S3

### Step 2 Analyse obstacles

Generally, goals are viewed idealistic and overlook unexpected behaviours of a real environment may cause their failures (van Lamsweerde and Letier 2000; Letier 2001). Taking a pessimistic view to goals, such situations, i.e. obstacles, should be systematically detected, assessed, and handled at the early stage of migration and if needed goals should be modified (Letier 2001). Obstacles are a dual notion of goals meaning that as goals capture desired conditions, obstacles capture undesirable conditions (Letier 2001). The framework defines an identify-assess-resolve cycle as follows:

(i) Identify obstacles that may impede satisfaction of goals (Step 2.1);
(ii) Assess risk of identified obstacles in terms of likelihood and criticality (Step 2.2); and
(iii) Resolve obstacles by modifying existing goals or generating new ones so as to prevent, reduce, or mitigate the obstacles (Step 2.3).

*Step 2.1 Identify obstacles*

The system architect can identify obstacles in two ways:

*(i)* *Evidential* where the probable obstacles are identified from the repository. For each goal in a goal model, the system architect reviews the collection of obstacles and shortlists probable ones. The shortlisting of obstacles is based on information provided by developers, user experience, statistics about systems, and available accounts about cloud services.

*(ii)* *Domain-based* where the obstacle is domain/platform specific and in fact is a refinement of an existing obstacle in the repository. Domain-specific obstacles are means to refine the goal model to new sub-obstacles.

Similar to goal elements, a parent obstacle might be a combination of other obstacles causing the parent obstacle (Letier 2001). Figure 4 shows goal *Achieve [Reduced data uploading time]* is obstructed by the obstacles *Performance variability of Amazon S3 (O27)* and *Geographical distance (O28)*. The root obstacle *Performance variability of Amazon S3 (O27)*, which is suggested by the repository, is refined into two domain-specific obstacles *High uploading time for blobs datatype (100k entries) (O27_1)* and *Low throughput to write buckets (O27_2)*. Moreover, the goal *Achieve [Improved availability]* is obstructed by the obstacles *Service transient fault (O3)* and *Cloud outage (O1)*. The obstacle *Cloud outage (O1)* suggested by the repository, by itself, is refined into three sub-obstacles *Local network disruption (O1_1)*, *I/O issues of servers (O1_2)*, and *S3 data centre outage (O1_3)* that are domain-specific refinements of the obstacle *Cloud outage (O1)*. The domain-specific obstacle *S3 data centre outage (O1_3)* is also refined into two obstacles *Local electrical storm (O1_3_1)* and *S3 power outage (O1_3_2)*.

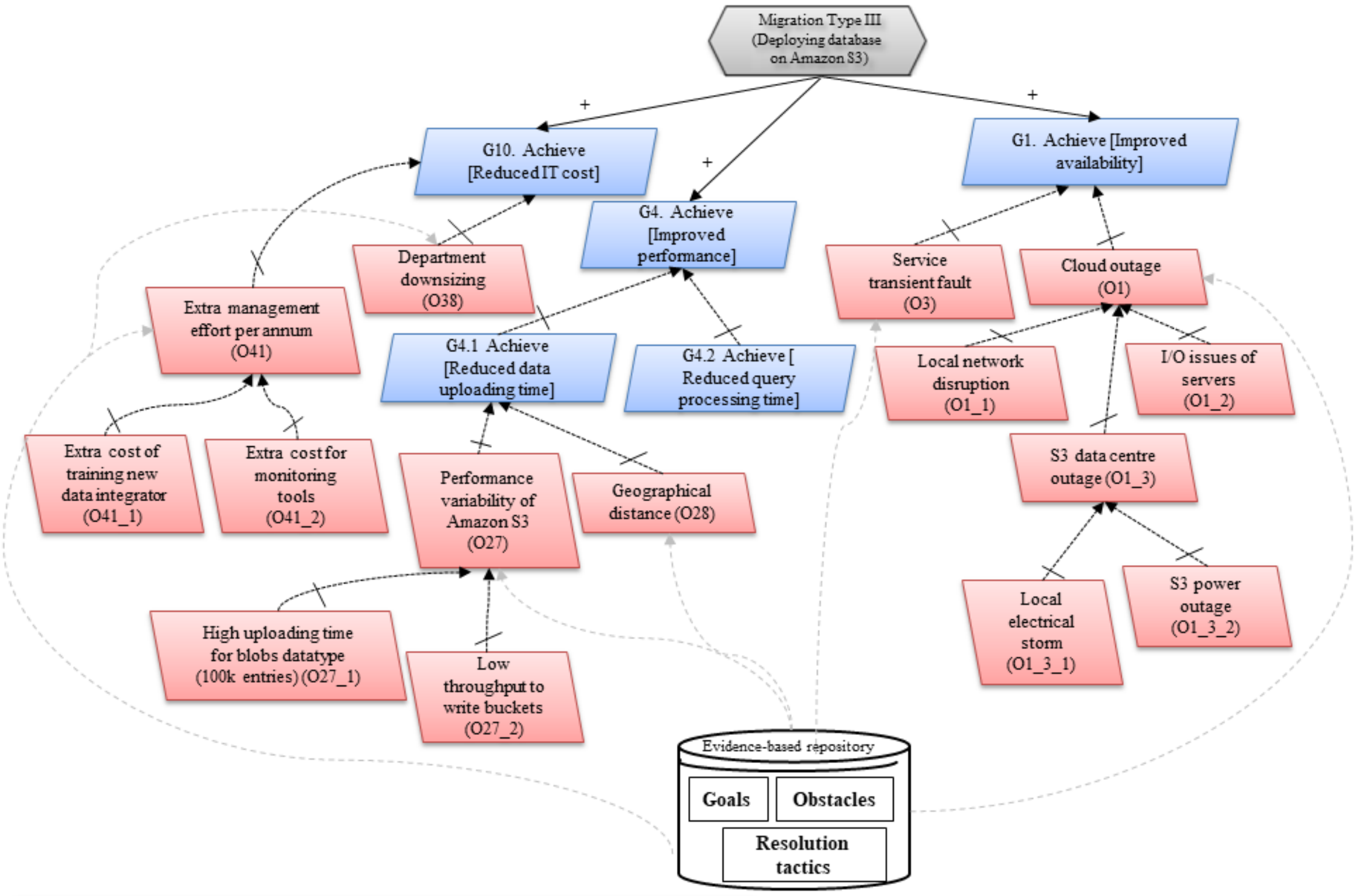

Figure 4. Obstacles against achieving quality goals *Achieve [Reduced IT cost]*, *Achieve [Improved performance]*, and *Achieve [Improved availability]* retrieved from the repository

### *Step 2.2 Assess Obstacles*

Analysing the risk or criticality of obstacles identified in Step 2.1 is important to get an understanding of requirements for making a legacy system cloud-enabled. The framework borrows a standard qualitative technique called Risk Analysis Matrix (RAM) devised by the acquisition reengineering team at the Air Force Electronic System Centre (Franklin 1996). The qualitative expression of obstacle risks in RAM is suitable if precise numerical techniques are difficult to find or not required. In RAM, the likelihood of an obstacle is judged by qualitative scales from *Almost Certain*, *Likely*, *Possible*, *Unlikely*, and *Rare* and the consequence of the obstacle occurrence is represented by *Insignificant*, *Minor*, *Moderate*, *Major*, and *Catastrophic*. These qualitative scales measure the likelihood of an obstacle occurrence and its associated consequences. The risk of an obstacle is defined as the product of its probability of occurrence and severity, i.e. Risk = Likelihood × Consequences. A risk matrix can be created to highlight the risk zone as shown in Table 4. An organization may define zones as generally unacceptable, acceptable, or low-risk. For example, the risk of an obstacle might be perceived as moderate (M) but it is still tolerable whilst an obstacle with H (High) and E (Extreme) should be handled more carefully.

Note that, calculating the product of likelihood and consequence of obstacles in Table 4 relies on the availability of information sources such as the specification of cloud services, statistics from legacy systems, developers, end-users' experience, and an overall impact of risks on goals. The system architect may use a voting mechanism involving stakeholders to accurately estimate the occurrence likelihood and consequences of obstacles. Hence, the values in Table 4 are actually computed based on the domain information in a goal-obstacle analysis scenario.

Table 4. risk matrix for obstacles

| | Consequence severity | | | | |
|---|---|---|---|---|---|
| Likelihood | Insignificant | Minor | Moderate | Major | Catastrophic |
| Almost Certain | H | H | E | E | V |
| Likely | M | H | H | E | V |
| Possible | L | M | H | E | E |
| Unlikely | L | L | M | H | E |
| Rare | L | L | M | H | H |

V: Very extreme risk, E: Extreme risk, H: High risk, M: Moderate risk, and L: Low risk

### *Step 2.3 Resolve Goal Obstacles*

Obstacles whose risks are recognized serious enough, (e.g. very extreme, extreme, and high risk) must be tackled. The framework relies on the repository's catalogue of resolution tactics to address obstacles identified in the previous step. In our framework, the tactics are cloud platform agnostic and vary among seven categories: namely *Goal/Service/Migration type Substitution*, *Obstacle prevention*, *Obstacle reduction*, *Goal weakening*, *Goal restoration*, *Goal mitigation*, and *Do nothing*. Their full definitions are presented in Appendix B. Resolution tactics are platform agnostic to give system developers freedom to evaluate a broad range of techniques to operationalize them. In this example, all the obstacles are deemed severe and thus the goal model is further refined down to resolution tactics (Figure 5). For instance, to reduce the occurrence likelihood of obstacle *Geographical distance (O28)*, the system architect chooses the resolution tactics *Refine network topology (T24)* from the repository. Another example is the reducing the risk of obstacles *High uploading time for blobs (100k entries) (O27_1)* and *Low throughput to write buckets (O27_2)* through incorporating the resolution tactic *Use multiple cloud servers (T27)* in the new cloud-enabled architecture of the system.

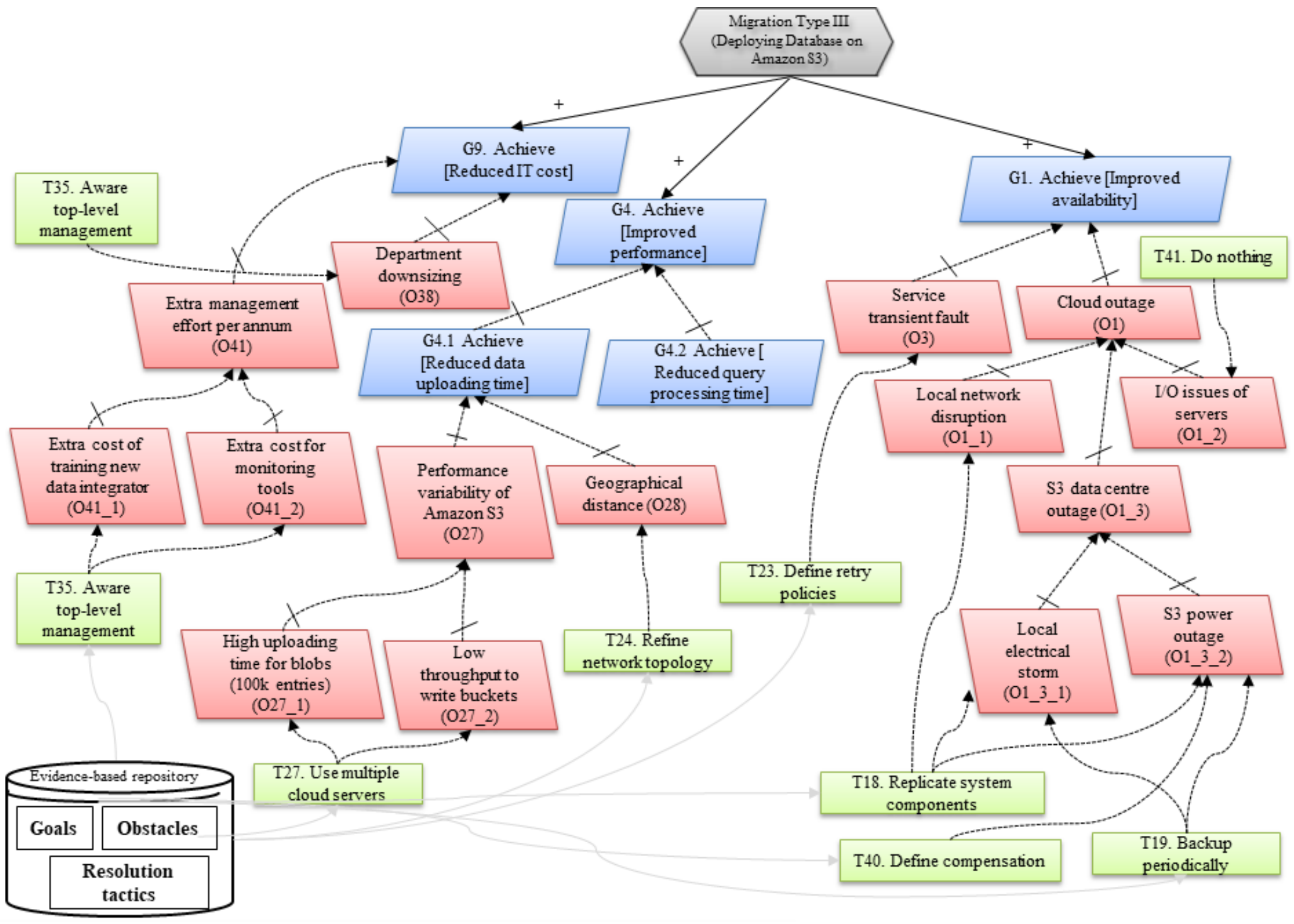

Figure 5. Resolution tactics to tackle obstacles

## 5. Application of the framework in practice

This section presents two case studies as a benchmark for validating the framework. They are instances of migration types V and IV. The first case is a scenario of moving an open-source Web-based system providing real-time stock quotes for users to a private cloud platform. In the second scenario, a Web-based system for processing digital documents is moved to a public cloud. The system architect uses domain information related to the scenarios to select and shortlist the pre-constructed collection of obstacles and resolution tactics. In both scenarios, the risk matrix values presented in Table 4 are used to assess obstacle risk.

### 5.1 Case study 1

*SpringTrader* is an open-source Web-based system that has been developed using J2EE framework and maintained by many contributor developers over time (Gordon 2015). Its architecture includes (i) a Web-based layer allows users creating an account, browsing stock portfolios, lookup stock quotes, and ordering stock trade orders and (ii) a backend that fulfils orders. The communication between the Web-based frontend and the backend is a-synchronous where the front-end delivers orders to a message queue and the back-end processes them.

Moving *SpringTrader* to Pivotal Cloud Foundry, that is an open source platform for developing and deploying full stack software systems in the cloud, enables users to access real-time stock market data in a more interactive way with the system as well as the individual scaling up/down of system components. The system architect analyses architectural requirements in enabling *SpringTrader* to operate in Cloud Foundry platform. The documentation of this project is available at (Gordon 2015).

#### *Step 1 Specify cloud migration goals*

A goal model is created with the three initial goals *Achieve [Increased scalability]*, *Achieve [Keeping system interoperable]*, and *Achieve [Keeping system available]* selected from the repository with the following specifications:

**Goal** Achieve [Increased scalability]
**Definition** [Moving the *SpringTrader* to the cloud should make it scalable in the sense that the system will be able to service massive end users' requests during workload]

**Goal** Achieve [Keeping system interoperable]
**Definition** [The *SpringTrader* should be integratable with and be able to call cloud services]

**Goal** Achieve [Keeping system availability]
**Definition** [Moving the *SpringTrader* to the cloud should not affect the system availability to end users]

### *Step 2 Analyse obstacles*

**Step 2.1 Identify obstacles**. Reviewing the architecture model of *SpringTrader* reveals that the tight dependencies among system component impede their individual scalability and portability across multiple instances of servers. This is an instance of the obstacle *Tight dependencies (O51)* against the goal *Achieve [Increased scalability]*. For the new platform, it is planned to use cloud database solutions MySQL and MongoDB for the *SpringTrader*. However, they are incompatible with the SQL database of *SpringTrader*. This is indeed an obstacle to the goal *Achieve [Keeping system interoperable]*, an instantiation of the root obstacle *Incompatibility of legacy data storage and cloud (O49)* defined in the repository. This obstacle, by itself, may occur in the form of two obstacles *Incompatible data operations (O50)* and *Incompatible data types (O21)*. Another obstacle to the goal *Achieve [Keeping system interoperable]* is that *SpringTrader* has been implemented using Java Development Kit 6 and Spring 3 that accordingly are not compatible with their equivalent (i.e. Java Development Kit 8 and Spring 4) in the Cloud Foundry platform. Integrating the *SpringTrader* with the Quote Web-Service, a service using the public Yahoo Finance APIs to provide real-time market data, may raise the risk of service unavailability as this service is hosted on the Cloud Foundry servers and geographically out of the local network of *SpringTrader*. This domain information confirmed that the obstacle *Service transient fault (O3)* is likely to occur against the goal *Achieve [Keeping system availability]*. The goal model is updated with new four obstacles shown in Figure 6.

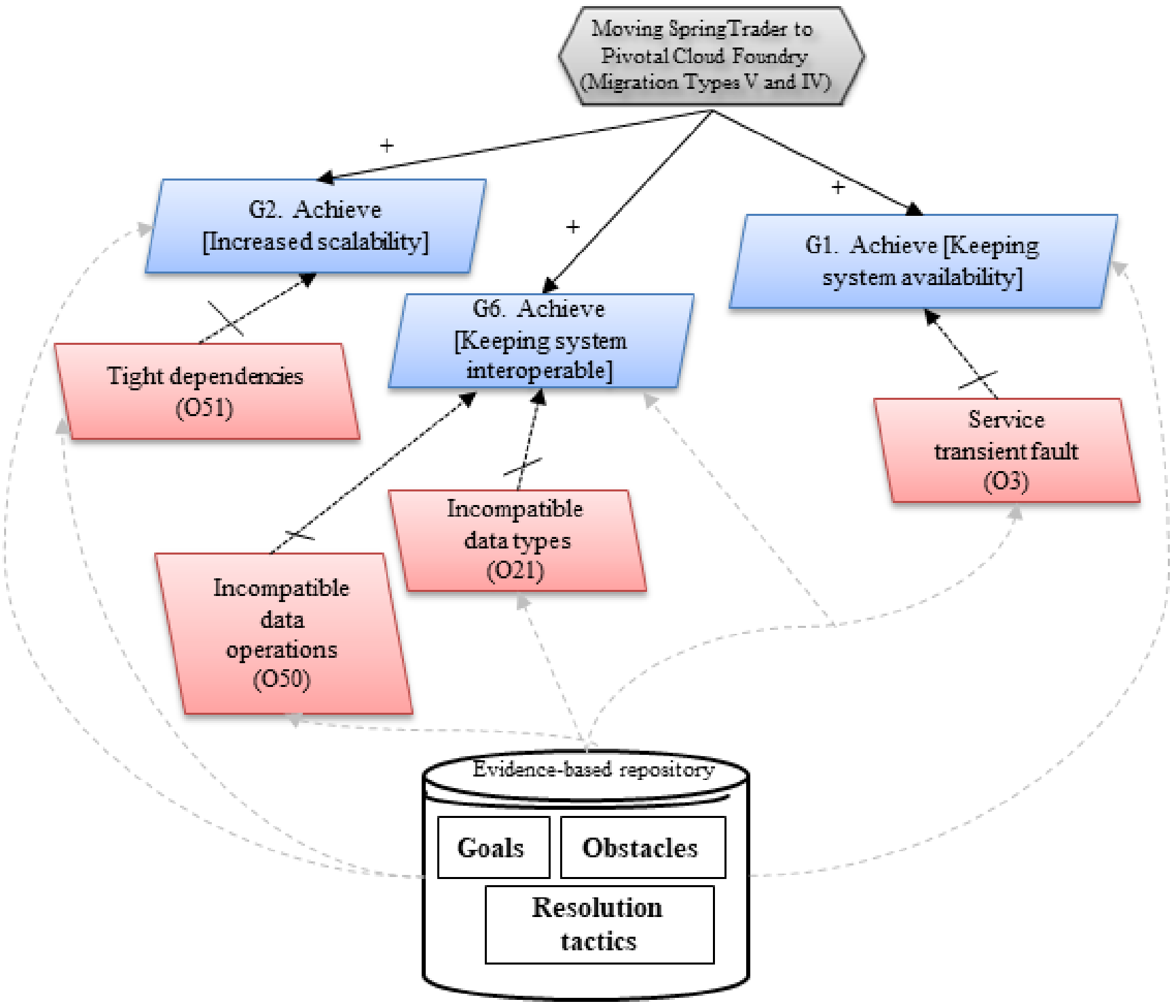

Figure 6. Goals *Achieve [Increased scalability]*, *Achieve [Keeping system interoperable]*, and *Achieve [Keeping system availability]* refined to four obstacles informed by the framework repository

**Step 2.2 Assess obstacles.** The occurrence probability and the consequence of the obstacles that identified from step 2.1 were assessed. Table 5 shows the risk matrix of obstacles. The goal *Achieve [Increased scalability]* is refined to the obstacle *Tight dependencies (O51)* as shown in Figure 7. According to the domain information, the system architect recognizes that the occurrence likelihood of this obstacle is *Almost Certain*. This is also true for the obstacles *Incompatible data operations (O50)* and *Incompatible data types (O21)* since *SpringTrader* database is incompatible with the Pivotal Cloud Foundry platform.

From past experience, developers believe that in some cases the occurrence likelihood of obstacle *Service transient fault (O3)* is *Possible* as *SpringTrader* components may not successfully call Quote Web-Service in the first attempt due to transient faults in making network connection to Quote Web-Service hosted in servers in Pivotal Cloud Foundry platform. Although it is a violation from the goal *Achieve [Keeping system available]*, its consequence is believed *Minor*. Therefore, the risk of this obstacle is set *Low*.

Table 5. Risk matrix for the obstacles identified from Step 2.1

| Obstacle against quality goal | Likelihood | Consequence | Risk |
|---|---|---|---|
| *Tight dependencies (O51)*. *SpringTrader* components have tight dependencies to meta-libraries that are sometimes incompatible with JDK 8. This cause the component of the system cannot be scalable and portable across multiple instances of servers. | Almost Certain | Major | E |
| *Incompatible data operations (O50)*. Various SQL statements in *SpringTrader* related to manipulating records are not syntactically and semantically compatible with corresponding MongoDB statements and MySQL provided by Pivotal Cloud Foundry platform. | Almost Certain | Major | E |
| *Incompatible data types (O21)*. Some data types (e.g. length and format) used in *SpringTrader* database are not compatible with corresponding ones in MySQL and MongoDB. | Almost Certain | Major | E |
| *Service transient fault (O3)*. Quote Web-Service might be temporarily unavailable due to network traffic or server workload. | Possible | Minor | L |

**Step 2.3 Resolve goal obstacles.** The system architect explored the repository to find resolution tactics that should be considered in new architecture of *SpringTrader* to operate in Pivotal Cloud Foundry platform. The system architect selectes the resolution tactic *Decouple system components (T7)* from the category *Obstacle prevention* to remove obstacle *Tight dependencies (O51)*. To operationalize the tactic a mediator and synchronisation mechanism is implemented to manage interaction between the system's components each deployed in different servers of Pivotal Cloud Foundry platform. For the obstacles *Incompatible data operations (O50)* and *Incompatible data types (O21)* the architect select the tactics *Adapt data (T12)* and *Develop adaptor/wrapper (T6)*, respectively. The former is to convert data types of *SpringTrader* into the data type of database solutions, i.e. MySQL and MongoDB, offered by the Pivotal Cloud Foundry platform whilst the latter is to add adaptors/wrappers that are responsible for runtime conversion of *SpringTrader* operations into the Pivotal Cloud Foundry.

To reduce the probability occurrence of the obstacle *Service transient fault (O3)*, the adopted resolution tactic is *Define retry policies (T23)* which is subsumed under the group *Goal Restoration*. That is, a retry policy is implemented in the architecture of *SpringTrader* to specify the required delay before executing the next attempt for connecting to the Pivotal Cloud Foundry server when transient faults occur due to network congestion. In addition, the system architect chooses the tactic *Replicate system components (T18)* from the group *Obstacle Prevention* group. The tactic is to partition, replicate, and distribute components/date (replicas) of *SpringTrader* over multiple servers of Pivotal Cloud Foundry.

Resolution tactics defined in the framework repository are generic recurrent solutions that can be operationalized using different implementation techniques or tools available in the cloud computing marketplace. In this scenario the resolution tactic *Develop adaptor/wrapper (T6)*, addressing incompatibilities between a system database and a cloud database solution, is operationalized using the notion of bounded context (Thönes 2015) in the sense that the transition of data is packed and unpacked during the executing of transactions. To realize the tactic *Decouple system components (T7)*, developers use *micro-service architecture design* (Dragoni, Giallorenzo, Lafuente et al. 2016) along with a service discovery mechanism to enable *SpringTrader* to locate micro services by name at a known catalogue endpoint and look them up dynamically at runtime. Figure 7 shows the resolution tactics selected.

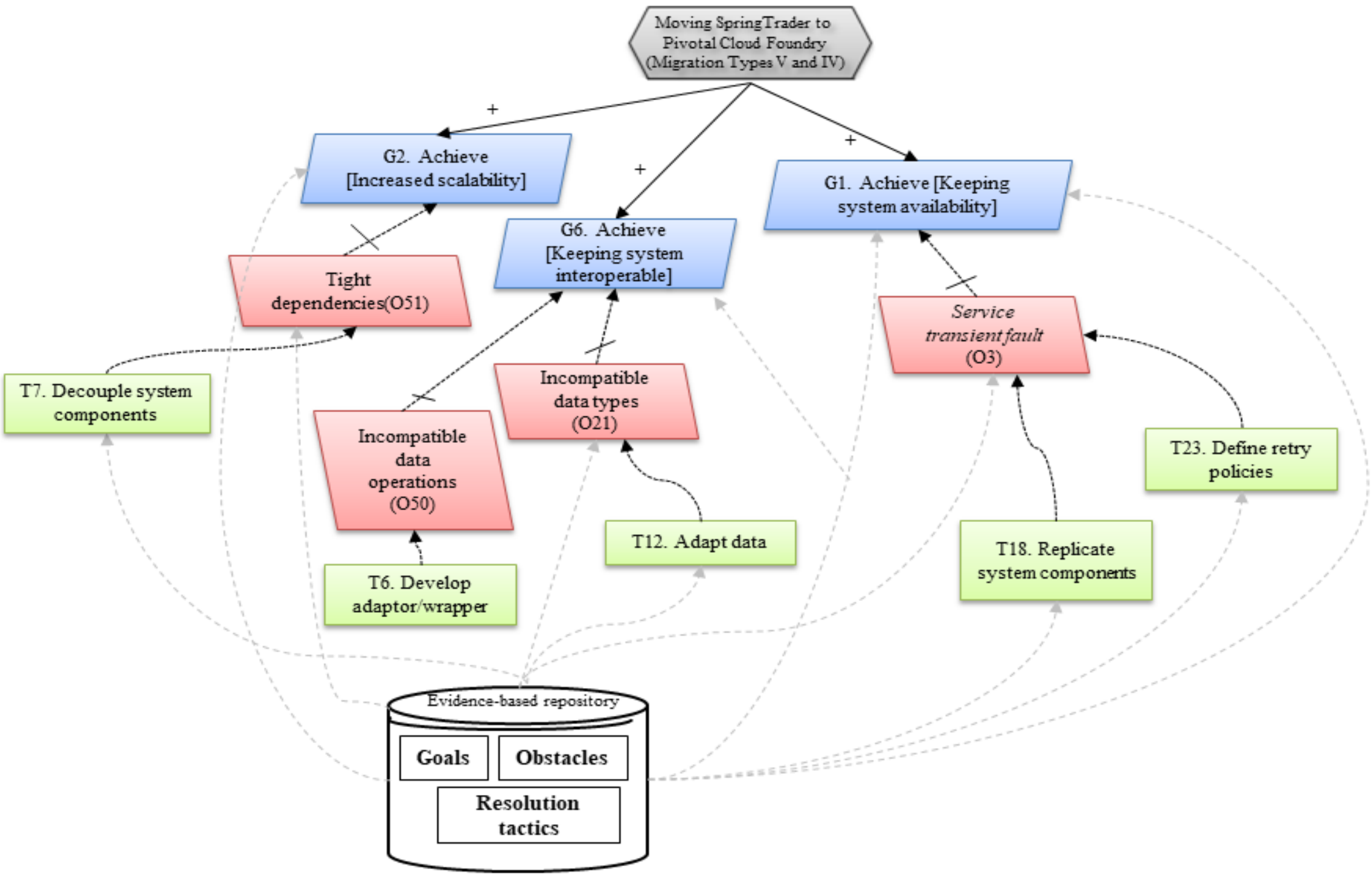

Figure 7.Resolutions tactics for handling obstacles

The first and second columns of Table 6, respectively, show the tactics and their operationalisation techniques used to handle the obstacles presented in the third column.

Table 6. Resolution tactics to handle obstacles in migrating *SpringTrader* to Pivotal Cloud Foundry

| Resolution tactic | Operationalisation | Relation to obstacle |
|---|---|---|
| Decouple system components (T7) | Decouple the *SpringTrader* components from each other by using mediator enabling a- synchronised interaction among loosely coupled components deployed on distributed architecture of Pivotal Cloud Foundry. | Tight dependencies (O51) |
| Develop adaptor/wrapper(T6) | Develop adaptor component in *SpringTrader* to emulate operations are supported in MySQL and MongoDB and map mismatches between datatypes in *SpringTrader* and Pivotal Cloud Foundry. | Incompatible data operations (O50) |
| Adapt data (T12) | Implement a mapping table to convert incompatible data types in *SpringTrader* and MySQL and MongoDB. | Incompatible data types (O21) |
| Replicate system components (T18) | Partition, replicate, and distribute the components of *SpringTrader* on multiple servers of Pivotal Cloud Foundry. | Service transient fault (O3) |
| Define retry policies (T23) | Implement retry policies in the source code of *SpringTrader* to specify the required delay before executing the next attempt when Pivotal Cloud Foundry does not respond. | |

### 5.2 Case study 2

The second case is adapted from the scenario presented in (Rabetski 2012; Rabetski and Schneider 2013). *InformIT* is a small independent software vendor in Sweden providing a Web-based digital document processing (DDP) system. The system offers publishing services to medium and large companies who own adequate infrastructure to perform these resource-demanding services. DDP is running on client companies' local infrastructure. Small companies are interested in taking the advantages of DDP's services. However, they cannot afford the financial commitment to procure new infrastructure, charging per user, and installation to use DDP. Small companies prefer to use DDP's services inconstantly and pay only for the amount of document processing. *InformIT* believes that DDP's services can be also used by small companies without the need for upgrading infrastructure if they are deployed in the cloud via migration types V and IV. The early stage goal-obstacle analysis conducts by the system architect regarding reengineering DDP to the cloud is described in the following.

***Step 1 Specify cloud adoption goals***

The cloud enablement scenario should not exceed 90 days. This is represented via the goal *Achieve [Reduced cloud adoption cost]* and its specification is:

> **Goal** Achieve [Reduced cloud adoption cost]
> **Definition** [According to *InformIT* policy, the latest completion time for any new technology adoption in small companies should not exceed more than 90 days. In this scenario, moving the DDP to the cloud should be fulfilled with minimum development effort].

Moreover, goals *Achieve [Improved performance]*, *Achieve [Improved testability]*, and *Achieve [Improved portability]* were expected to be satisfied by moving DDP to a cloud platform. For example, the goal *Achieve [Improved performance]* is defined:

> **Goal** Achieve [Improved performance]
> **Category** Performance Goal
> **Definition** [acceptable system throughput for rendering a digital document with any size should be no more than 4.9 seconds].

***Step 2 Analyse obstacles***

**Step 2.1 Identify obstacles.** In the view of domain information, scanning the framework repository refines the top goals towards root obstacles and subsequently leaf ones (Figure 8). For example, there are two probable obstacles *Learning curve (O33)* and *Incompatibility of legacy and cloud service (O48)* against the satisfaction of the system goal *Achieve [Reduced cloud adoption cost]*. Moreover, experience of developers confirmed that the goal *Achieve [Improved performance]* might be obstructed by the performance variability of cloud servers once DDP is in operation on cloud servers. This is shown by the obstacle *Performance variability of cloud server (O27)* in the goal model (Figure 8).

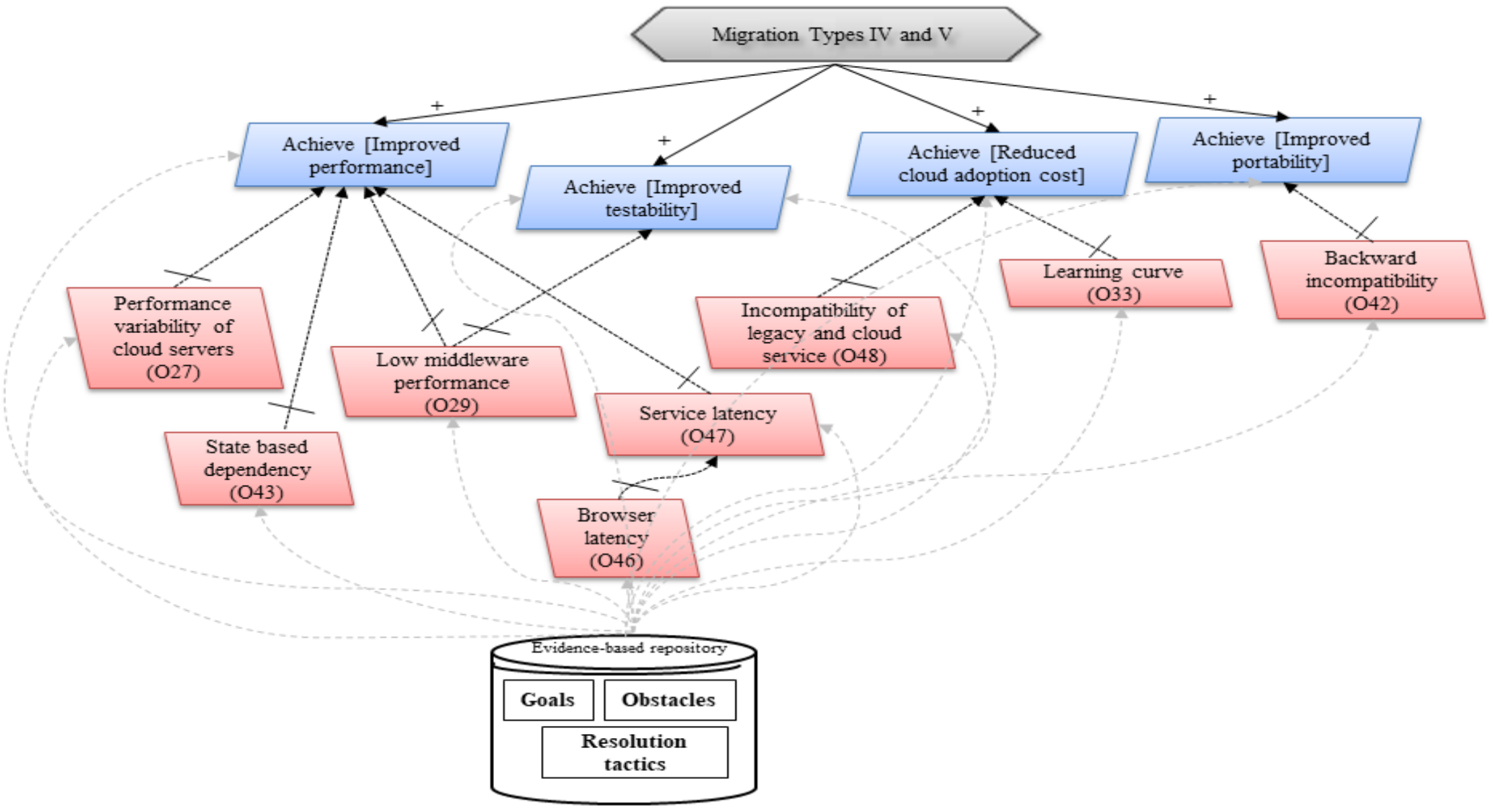

Figure 8 Goals *Achieve [Improved performance]*, *Achieve [Improved testability]*, *Achieve [Reduced cloud adoption cost]*, *Achieve [Improved portability]* refined to leaf obstacles

**Step 2.2 Assess obstacles**. Technical documents of DDP and an early investigation of public cloud platforms reveals the occurrence of obstacles *Learning curve (O33)* and *Incompatibility of legacy and cloud service (O48)* is *Almost Certain* with a *Major* consequence on the satisfaction of the goal *Achieve [Reduced cloud adoption cost]*, indicating an *Extreme* risk. All the leaf obstacles are assigned a risk value based on the likelihood of their occurrence and consequence as shown in Table 7.

Table 7. Risk matrix for the obstacles identified from Step 2.1

| Obstacle | Likelihood | Consequence | Risk |
|---|---|---|---|
| Performance variability of cloud servers (O27) | Likely | Major | E |
| State based dependency (O29) | Likely | Moderate | H |
| Low middleware performance (O29) | Likely | Moderate | H |
| Browser latency (O46) | Likely | Moderate | H |
| Incompatibility of legacy and cloud service (O48) | Almost Certain | Major | E |
| Learning curve (O33) | Almost Certain | Major | E |
| Backward incompatibility (O42) | Likely | Moderate | H |

**Step 2.3 Resolve goal obstacles.** The system architect tries to tackle obstacles *Learning curve (O33)* and *Incompatibility of legacy and cloud service (O48)* by using the resolution tactics *Substitute cloud service (T3)* and *Goal weakening (T36)*. *Substitute cloud service (T3)* is to select a cloud service/provider in a way that the new selected cloud service can still contribute to quality goals. As DDP has been developed with Microsoft family technologies and developers had programming experience of, choosing Microsoft Azure cloud platform is taken precedence over other popular cloud platforms such as Amazon Web Service and Google App Engine. This choice can also contribute to the goal *Achieve [Reduced cloud adoption cost]* by decreasing the likelihood occurrence of incompatibilities between DDP and Microsoft Azure cloud platform from initial value *Almost Certain* to *Possible*.

In some cases that an obstructed goal is found to be very idealistic, its definition can be changed to make its constraints relaxing in a way that the obstruction occurrence becomes tolerable. In this regard, the tactic *Degrade goal (T36)* is used by for the goal *Achieve [Reduced cloud adoption cost]* by extending the project deadline from 90 to 120 days. Figure 9 shows the produced goal model thus far.

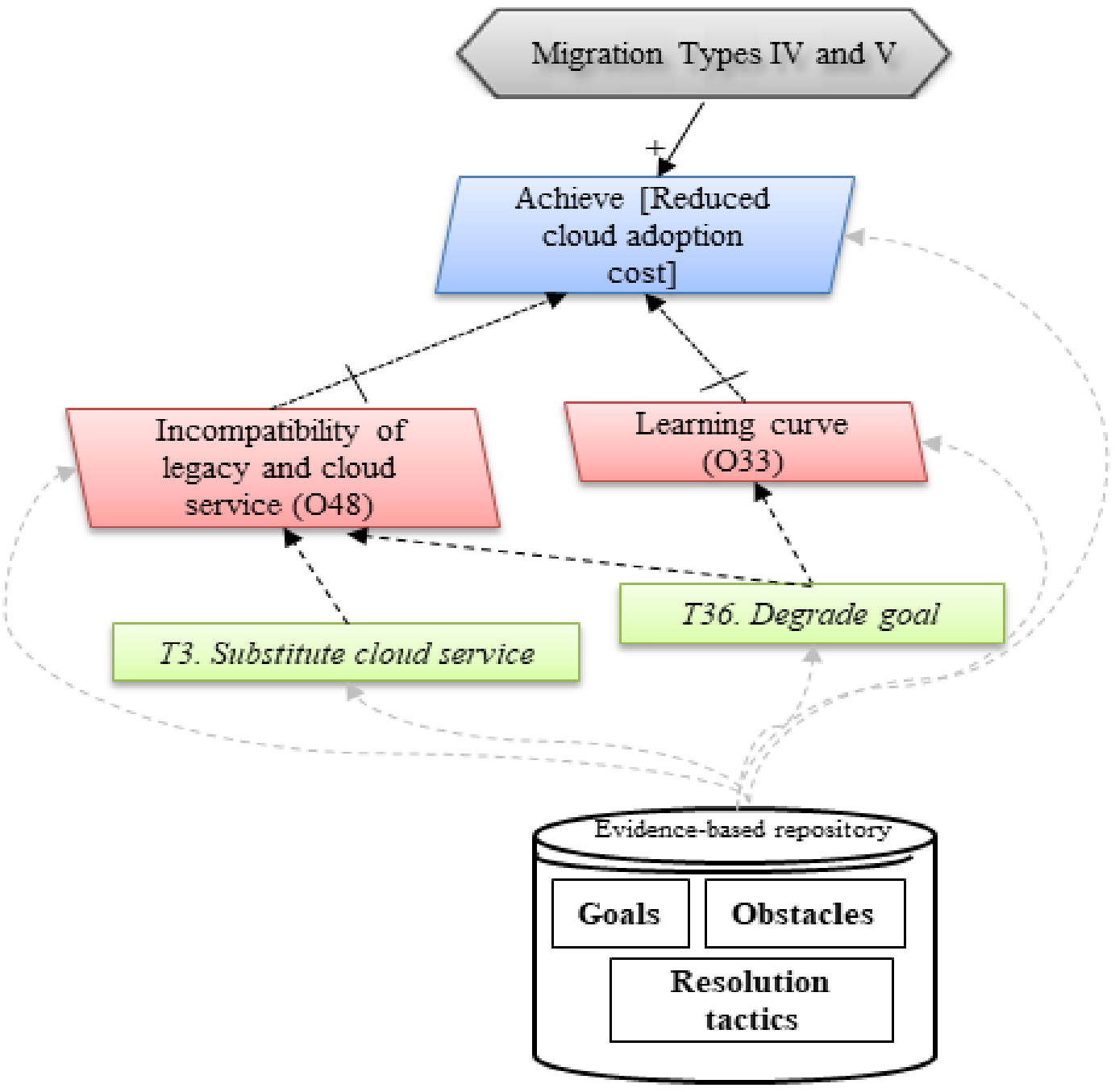

Figure 9 Obstacles to the goal *Achieve [Reduced cloud adoption cost]* and applied resolution tactics *Substitute cloud service (T3)* and *Degrade goal (T36)*

The obstacle resolution is an iterative process in the sense that once a tactic is chosen, it may raise new obstacles that should be resolved accordingly by reiterating steps 2.1 to 2.3 and refining the goal model. In the current scenario, despite applying the resolution tactic *Substitute cloud service (T3)* to reduce the obstacle *Incompatibility of legacy and cloud service (O48)*, the domain information about DDP and cloud platform Microsoft Azure documentation confirms that the violation of the goal *Achieve [Reduced cloud adoption cost]* is still possible because DDP's APIs are not compatible with their counterparts in the Microsoft Azure cloud platform. The obstacle *Incompatibility of legacy and cloud service (O48)* is refined into two obstacles *Incompatible APIs (O44)* (i.e. between DDP and Microsoft Azure platform) and *Incompatibility of legacy data storage and cloud (O49)*. The parent obstacle *Incompatibility of legacy data storage and cloud (O49)* is also split into two leaf domain specific obstacles (Figure 10). The definition of the leaf obstacles against the goal *Achieve [Reduced adoption cost]* is as follows:

**Obstacle** *Incompatible APIs (O44)*
**Definition** [DDP uses API's offered by .NET 2.0 and Visual Studio 2005 which might not be compatible with Microsoft Azure platforms].

**Obstacle** *Incompatible datatypes (O21)*
**Definition** [Datatypes in DDP are based on SQL Server Database .NET 2.0 platform which might not be compatible with Microsoft Azure database solution].

**Obstacle** *Incompatible data operations (O50)*
**Definition** [Data operations in DDP supported by SQL Server Database .NET 2.0 platform might not be compatible with Microsoft Azure database solution].

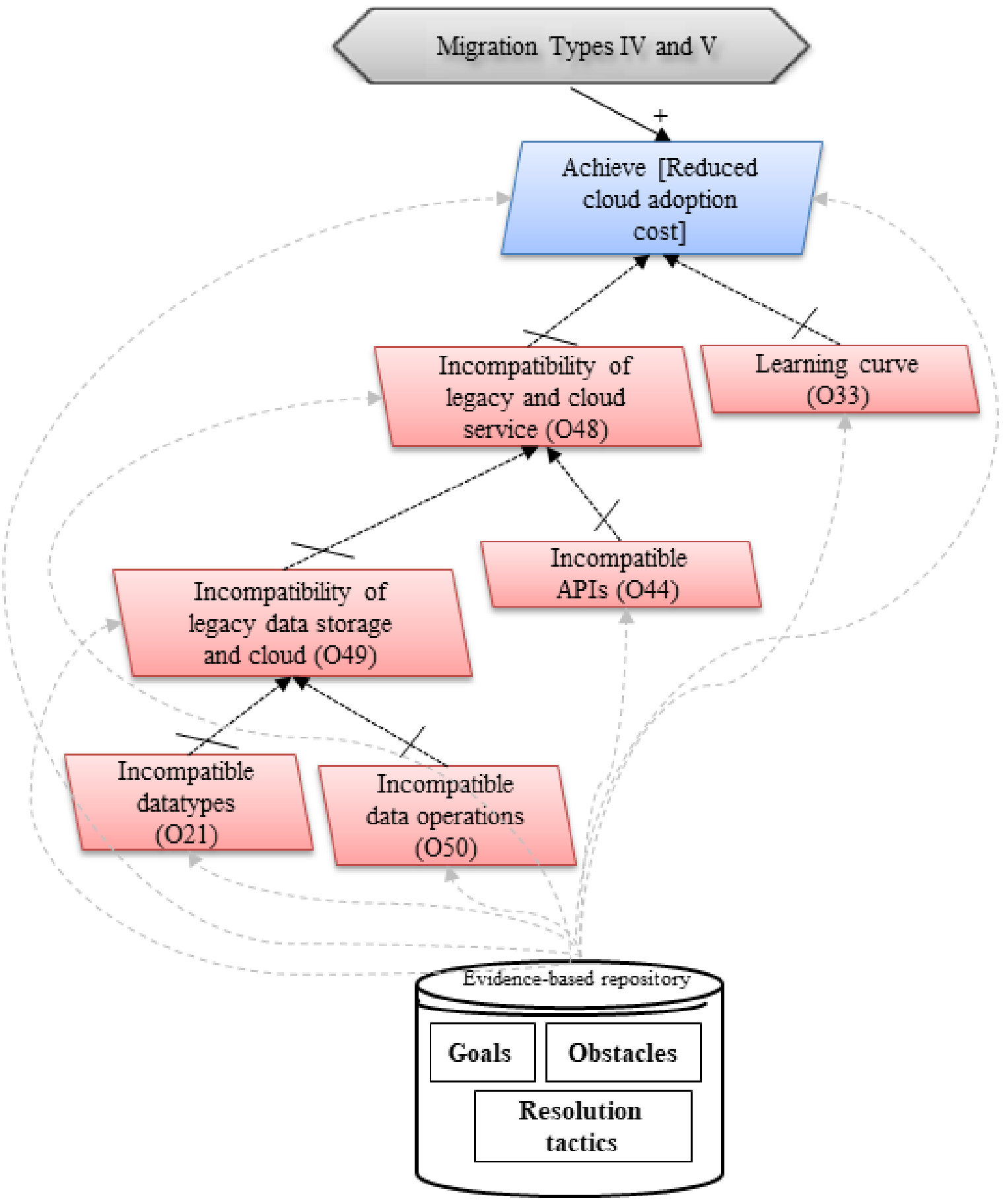

Figure 10 Refinement of goal *Achieve [Reduced IT cost]* to obstacles

In addition, applying tactic *Substitute cloud service (T3)* generated new obstacles specific to Microsoft Azure Platform. That is, the root obstacle *Microsoft Azure middleware latency (O29)* is refined into three leaf obstacles, *Microsoft Azure database middleware latency (O29_1)*, *Microsoft Azure message middleware latency (O29_2)*, and *Microsoft Azure transaction middleware latency (O29_3)*. Furthermore, the obstacle *Service latency (O47)* is decomposed into two obstacles *On-premise hardware latency (O47_1)* and *Distance from Microsoft Azure servers (O28)*. Figure 11 shows goal model refined after using the resolution tactic *Substitute cloud service (T3)* and based on evidential information provided from the repository. For simplicity, the system architect changes the original names of some obstacles identified from the repository but obstacle codes left unchanged. For example, in Figure 11, the obstacle *Backward incompatibility (O42)* is changed to *Switch between regular file system API to Microsoft Azure Storage API (O42)*.

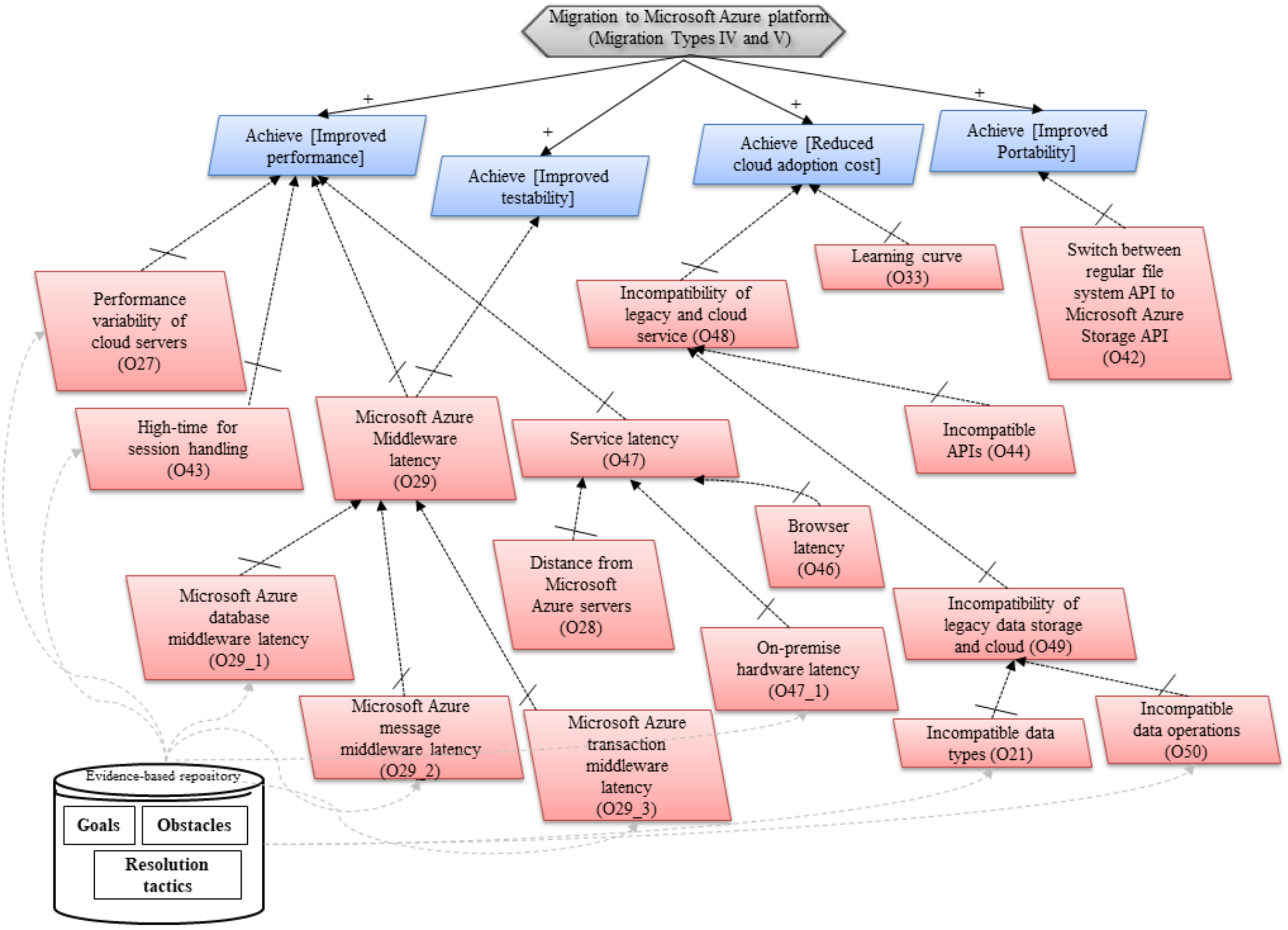

Figure 11 Refined goals models after identifying obstacles

To handle new obstacles generated as a result of applying the tactic *Substitute cloud service (T3)*, the system architect selects resolution tactics from the category *Obstacle prevention*. For the obstacles *Incompatible data operations (O50)* and *Incompatible datatypes (O21)*, the system architect uses *Develop adaptor (T6)* and *Adapt data (T12)*, respectively. The former is to implement a wrapper component which hides incompatibilities (e.g. queries and stored procedures) between the data layer of DDP and Microsoft Azure SQL whilst the later tactic is to convert SQL data types used in DDP to Microsoft Azure SQL database. To resolve the obstacle *Incompatible APIs (O44)*, the tactic *Develop adaptor (T6)* is used. For the obstacle *High-time for session handling (O43)*, the tactics *Make system stateless (T29)* is applied. Also, in handling the obstacle *Switch between regular file system API to Microsoft Azure Storage API (O42)*, the architect picks the tactic *Decouple system components (T7)* from the repository.

To reduce the probability occurrence of the root obstacle *Microsoft Azure middleware latency (O29)*, the adopted resolution tactic is *Refine network topology (T24)* which belongs to *Obstacle reduction* group. This tactic is operationalized through selecting Microsoft Azure servers close to *InformIT*'s network located in North Europe. For the obstacle *Browser latency (O46)* the tactic *Update patches (T21)* is used regularly.

Furthermore, in addressing the obstacle *Performance variability of Microsoft Azure servers (O27)*, the system architect applies *Degrade goal (T36)*, a tactic from *Goal weakening* group, to modify the definition of satisfaction level for the root goal *Achieve [Improved performance]*. The tactic refines this goal to a more liberal one via allowing the expected processing time of documents by DDP to be varied up to 2 hours in a peak time for documents with size more than 40 megabytes. The suggested tactic is hard to get acceptance by DDP users; however, the purpose of considering this tactic is to probe possible solutions to tackle the obstacle. Hence, the second tactic is *Acquire more cloud resources (T26)* operationalized by adding 3 more virtual servers.

The occurrence likelihood of the obstacle *On-premise hardware latency (O47_1)* was found *Possible* with a consequence as *Insignificant* (i.e. Possible * Insignificant = Low risk) and thus left unresolved. This is an instance of tactic *Do-Nothing*. Figure 12 shows the resultant goal model and incorporation of resolution tactics into the system architecture. Table 8 summarises all identified obstacles and selected resolution tactics.

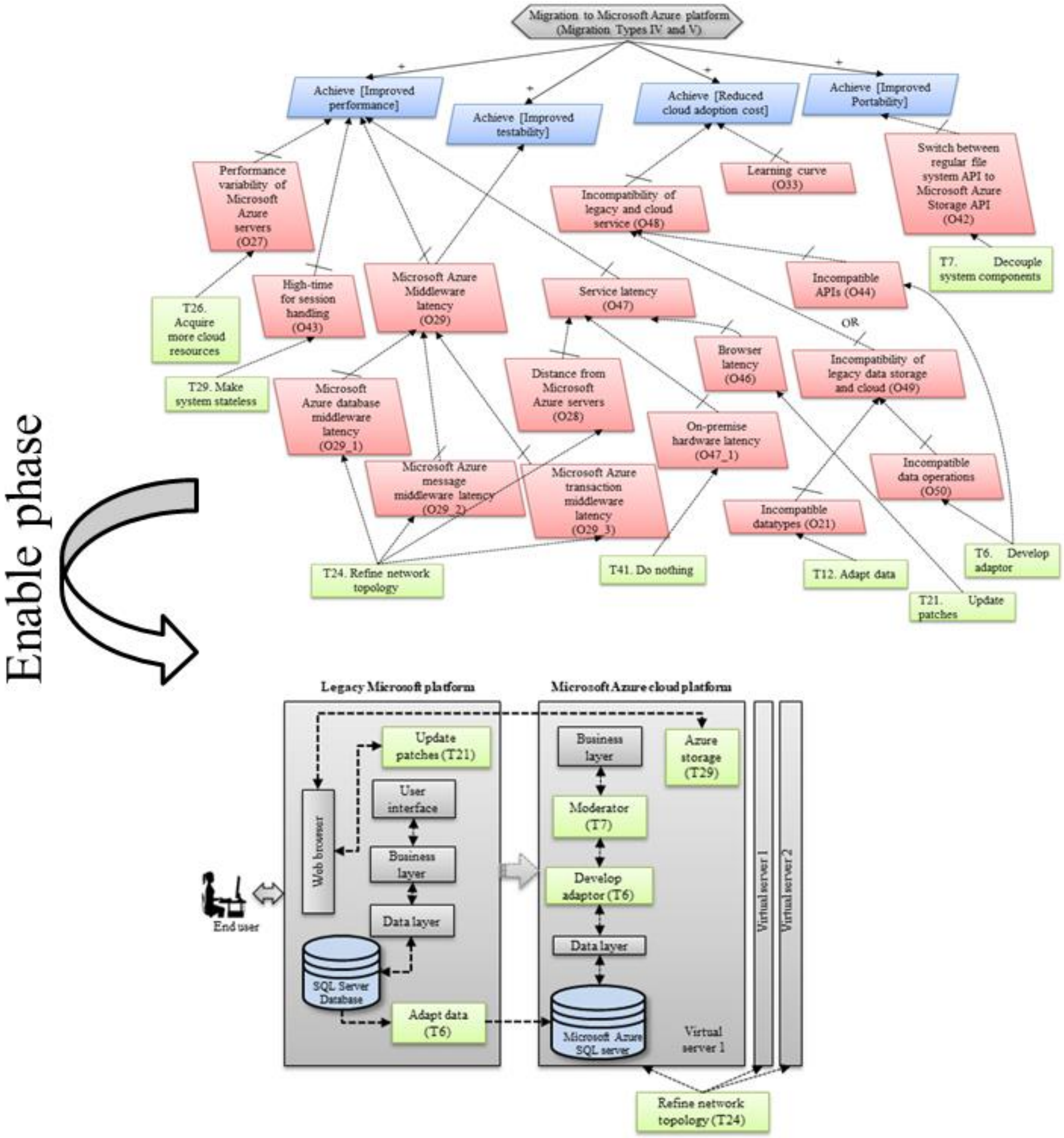

Figure 12 Resolutions tactics for handling obstacles are incorporated into a new architecture of DDP during the phase *Enable*

Table 8 Goals, obstacles and their risk, and selected resolution tactics

| Goal | Obstacle | | | | Adopted resolution tactics |
|---|---|---|---|---|---|
| | Obstacle id | Likelihood | Consequence | Risk | |
| Achieve [Reduced adoption cost] | O33 | Almost Certain | Major | E | *Substitute cloud service (T3).* Among three major cloud platforms Amazon Web Services, Google App Engine, and Microsoft Azure, the system architect choses Microsoft Azure because the legacy system has been developed using Microsoft family technology and developers has consistent experience with it. This reduces the cost of learning cloud technology and also potential effort in addressing incompatibilities between these platforms. |
| | O44 | Likely | Moderate | H | *Develop adaptors (T6).* A wrapper component is developed to resolve API mismatches between legacy system and Microsoft Azure. It hides specific Microsoft Azure characteristics that cause conflicts with the legacy system. |
| | O21 | Almost Certain | Moderate | E | *Adapt data (T12).* The legacy system data types are converted to Microsoft Azure cloud database solution. |
| | O50 | Almost Certain | Moderate | E | *Develop adaptor (T6).* As the Microsoft Azure does not support some kind of stored procedures, an emulator is implemented and deployed in Microsoft Azure server which performs missing functionalities that are not supported by this platform. |
| Achieve [Improved Portability] | O42 | Likely | Moderate | H | *Decouple system components (T7).* Legacy system components are decoupled so that dependency among them is minimized and they can work independently and interact in a-synchronised way. A mediator component is implemented to manage interaction between the loosely coupled components deployed on Microsoft Azure servers. |
| Achieve [Improved testability] | O29_2 | Likely | Moderate | H | *Refine network topology (T24).* Define the geographical location of virtual machines close to North Europe to minimize latency of Azure middleware. |
| | O29_1 | Likely | Moderate | H | |
| | O29_3 | Likely | Moderate | H | |
| Achieve [Improved performance] | O27 | Likely | Major | E | *Acquire more cloud resources (T26).* Rent three virtual machines to address slow CPU clock rates. Use physical disk shipping to reduce effects of network latency/transfer rates. Use third party monitoring tools to independently verify the system performance. |
| | O29_2 | Likely | Moderate | H | |
| | O29_1 | Likely | Moderate | H | |
| | O29_3 | Likely | Moderate | H | |
| | O29 | Likely | Moderate | H | *Make system stateless (T29).* Legacy system components should be modified in a way that they do not depend on internal state. Rather, such states should be stored in an external storage or requested from an external component. |
| | O28 | Likely | Moderate | H | *Refine network topology (T24).* Modify the current deployment and distribution model of legacy system components on the basis of transaction delay, proximity, and geographical distribution. In this case, system components are deployed in Microsoft Azure servers located in North Europe close to Sweden (the location of the system) to reduce latency. |
| | O46 | Likely | Moderate | H | *Update patches (T21).* Update cloud service consumer browsers regularly. |
| | O47_1 | Possible | Insignificant | L | *Do nothing (T41).* This obstacle is not perceived critical. |

## 6. Related work

Early stage analysis of cloud computing adoption has been previously investigated by other authors. Drawing upon some relevant studies to compare decision making frameworks e.g. (Babar, Liming and Jeffery 2004) and following guidelines for legacy system cloud enablement (Fahmideh, Daneshgar, Low et al. 2016), eight analysis criteria were identified: migration type, lifecycle focus, evaluation granularity, evaluation approach, process support, stakeholder involvement, modelling language, experience repository, and tool support. In what follows, existing literature is discussed in view of these criteria. This will situate the proposed framework in the literature and highlight its contributions to the state of art.

**(i) migration type.** As mentioned in Section 4.1, there are few options, namely type I, II, III, IV, and V, through which legacy system benefit from cloud service. In view of this criterion, (Anstett, Leymann, Mietzner et al. 2009) presents some factors such as operating system and platform middleware that a system architect should take into account when deploying business process engines in IaaS, i.e. migration type I. For migration type II in which system components are substituted with cloud services through SaaS model, (Godse and Mulik 2009) and (Wu, Lan and Lee 2011) are two example frameworks presenting approaches for selecting the most appropriate SaaS product for organizational needs. For the migration type V where the whole system stack is encapsulated in virtual machines and then ran in the cloud infrastructure, the framework proposed in (Khajeh-Hosseini, Greenwood, Smith et al. 2012) supports decision makers in identifying concerns to examine the costs of deploying IT systems on the cloud. We did not find any noticeable framework in relation to migration types III and IV.

**(ii) lifecycle focus.** Decision making frameworks can be classified considering migration phases for which a framework is appropriate to use. Fahmideh et al. define three migration phases: plan, enable, and maintain (Fahmideh 2016b). Frameworks related to the planning phase are concerned with the feasibility assessment of adopting cloud services. Some studies such as (El-Gazzar, Hustad and Olsen 2016), (Low, Chen and Wu 2011) and (Wu 2011) mainly investigate important factors such as security, cost, and organizational readiness that should be taken into account when adopting cloud services. Other approaches proposed in (De Assunçao, Di Costanzo and Buyya 2009), (Deelman, Singh, Livny et al. 2008), (Kondo, Javadi, Malecot et al. 2009), (Walker, Brisken and Romney 2010), and (Khajeh-Hosseini, Greenwood, Smith et al. 2012) focus on analysing the feasibility of cloud adoption from a cost saving point of view. This informs if a deployment option is cost effective. Other concerns related to re-architecting systems to cloud such as data security, interoperability between legacy components and cloud services, or system performance are typically not covered in that analysis. In the enablement phase, cloud services that meet given computational requirements of systems and the most suitable system components that can benefit from cloud services are first selected. This is followed with defining an optimum deployment of the components in cloud servers. Multi-criteria based decision making techniques are applied for shortlisting and ranking of candidate cloud services: e.g. Analytic Hierarchy Process (AHP) (Garg, Versteeg and Buyya 2013), (Godse and Mulik 2009), and Analytic Network Process (ANP) (Menzel, Schönherr and Tai 2013). There are situations in which post-migration assessment is needed. The decision making frameworks related to this phase may suggest which system components are either de-migrated to the local environment or enhanced with new cloud services based on new needs and changes in the operational environment (Scandurra, Mongiello, Colucci et al. 2016).

**(iii) evaluation granularity.** The unit of analysis in decision making frameworks can be at different levels such as organizational, system, or system component. A system architect may proceed at one of these levels to evaluate the suitability and filter out migration variants that do not meet requirements. For example, some studies such as (Nikkhouy 2013), (Christoforou and Andreou 2013), and (Low, Chen and Wu 2011) assess whether an organisation is ready to benefit from cloud services. Other frameworks (Tak, Urgaonkar and Sivasubramaniam 2011), (Juan-Verdejo and Baars 2013), (Menzel and Ranjan 2012), (Khajeh-Hosseini, Greenwood, Smith et al. 2012), (Fittkau, Frey and Hasselbring 2012), (Saripalli and Pingali 2011), and (Calheiros, Ranjan, Beloglazov et al. 2011) examine which systems are adequate for moving to the cloud using migration types V. Furthermore, (Leymann,

Fehling, Mietzner et al. 2011) suggests an approach to identify system components suitable for being cloud-enabled based on other factors such as latency, data transfer, and component dependencies.

**(iv) evaluation approach.** This criterion is useful to know the level of information and technique are required for an evaluation exercise. Decision making frameworks may use a wide range of techniques to satisfy desired goals. To name a few, some frameworks are metric based such as (De Assunçao, Di Costanzo and Buyya 2009), (Deelman, Singh, Livny et al. 2008), and (Kondo, Javadi, Malecot et al. 2009), some use goal-based reasoning such as (Zardari, Bahsoon and Ekárt 2014) and (Scandurra, Mongiello, Colucci et al. 2016), some use optimisation technique such as (Leymann, Fehling, Mietzner et al. 2011) and other use hybrid techniques such as (Menzel and Ranjan 2012), (Fittkau, Frey and Hasselbring 2012), (Saripalli and Pingali 2011).

**(v) process support.** Decision making frameworks may define a precise definition of their steps and sequencing. They may clearly describe for each step the input and the output products, their guidelines, controls, and any heuristics for an accurate assessment. They ultimately help a system architect to accomplish decision making goals. Except for some frameworks presented in Table 9, the majority of frameworks reviewed in this section provide at least a general description of their process, though they differ in the level of details provided. However, it should be said that some frameworks related to the planning phase do not have an explicit process support.

**(vi) stakeholder involvement.** As with many decision making scenarios in software engineering, cloud computing adoption may involve multiple stakeholders such as cloud service providers, consumers, brokers, developers, project managers, and end-users whose interests attempt to influence risks and benefits. These stakeholders may well have their own competing interests and attempt to influence risks and benefits of an assessment process. Their active participation enables proper elicitation of their goals and priorities. Resolving conflicts during decision making process is essential for the quality of the assessment. Existing frameworks generally recognize the importance of incorporating key stakeholders, but they vary in how and to what degree stakeholders are engaged.

**(vii) modelling language.** Evaluation frameworks can be compared in terms of if and how they employ a notation to represent elements, semantic interpretation, and outcome of each decision step. Using a modelling language can facilitate communications and understandability of the decision making process to stakeholders. It can also provide a scope for automation. For example, the framework by (Christoforou and Andreou 2013) uses Influence Diagrams, a directed acyclic graph with nodes, to show decision variables and how they influence each other. Nodes representations along with their dependencies model decision making questions and provide final decision nodes. In another framework, (Zardari, Bahsoon and Ekárt 2014) goal-oriented modelling is used to represent risks encountered and mitigating strategies for using cloud services.

**(viii) experience repository.** The notion of reuse is a perennial means for increasing productivity in software engineering (A. Aurum 2003). Like any other software development activity, cloud migration decision making is a knowledge-intensive process. It can be a costly exercise if it starts from scratch in an ad-hoc manner each time. The effort involved can clearly be reduced if knowledge from activities in previous adoption scenarios is maintained and reused. Towards this, CLiCk (Cloud Life Cycle) provides a repository containing historical information on QoS of different service platforms to improve the accuracy of service selection (Giovanoli 2012). Reusing and sharing recurring decision logs in the course of re-architecting legacy systems to cloud platforms is suggested in (Zimmermann, Wegmann, Koziolek et al. 2015). In another work, (Menzel, Schönherr and Tai 2013) provides a reusable catalogue of criteria for creating customized evaluation methods to evaluate alternative service providers.

**(ix) tool support.** A decision making process can involve time-consuming tasks such as collecting, documenting, and maintaining relevant domain data. Particularly, at the early stage of transition to the cloud, consumers may face a higher number of risks in utilizing cloud services which should be carefully evaluated (Lacity, Khan and Willcocks 2009). Decision-making tools can capture, for example, alternative cloud services and their service level agreements offering by different providers, costs and risk factors relevant to decision scope, automate as many as decision steps, and come up with evaluation outcomes. The importance of tool support is recognized in some frameworks such as (Khajeh-Hosseini, Greenwood, Smith et al. 2012) and (Menzel and Ranjan 2012).

Table 9 summarizes characterizing the existing studies. Our proposed framework in the current study distinguishes itself from the existing one in the view of analysis criterion *experience repository*. Compared to the existing studies reviewed above, our framework provides an evidential knowledge repository of reusable cloud-specific obstacles and corresponding resolution tactics along with a visualization mechanism for systematically analysing risks in migrating systems to cloud platforms. Perhaps, the only notable close work to our framework is by (Giovanoli 2012) which provides a repository containing information on the different service providers and their services. However, it does not cover several areas of obstacles and resolution tactics (e.g. incompatibilities between systems and cloud platforms). None of the existing studies utilises cloud adoption knowledge during goal reasoning to address probable risks and to undertake countermeasures. They rather rely on knowledge of system architect which might be imprecise and incomplete. Using our framework, system architects can get an informed insight of attainability of system quality goals via cloud-enablement of systems. They also get a detailed goal-obstacle analysis supported by the evidential knowledge repository.

Table 9. Literature comparison addressing the evaluating of cloud computing adoption

| Study | Aim | Migration Type | Lifecycle Focus | Evaluation granularity | Evaluation approach | Process support | Stakeholder involvement | Modelling language | Experience repository | Tool Support |
|---|---|---|---|---|---|---|---|---|---|---|
| (Anstett, Leymann, Mietzner et al. 2009) | Identifying factors such as operating system, platform middleware and legacy system to be considered in deploying business process execution language (BPEL) on IaaS. | Type V | Plan phase | Legacy system | Not specified | Not specified | Not specified | Not specified | Not considered | Not available |
| (Khajeh-Hosseini, Greenwood, Smith et al. 2012) | Providing a decision making support for identifying concerns in using IaaS. | Type V | Plan phase | Legacy systems | Cost modelling | Yes | Yes | Deployment model of system | Not considered | Yes |
| (El-Gazzar, Hustad and Olsen 2016) | Identifying inhibitors and organisational drivers are involved in a decision making for cloud computing adoption. | All | Plan phase | Organisation | Not specified | Not specified | Not specified | Not specified | Not considered | Not available |
| (Low, Chen and Wu 2011) | Exploring factors affecting organisations in adopting cloud computing. | All | Plan phase | Organisation | Not specified | Not specified | Not specified | Not specified | Not considered | Not available |
| (Wu, Lan and Lee 2011) | Exploring factors influencing successful SaaS adoption. | Type II | Plan phase | Organisation | Decision making trial and evaluation laboratory | Not specified | Not specified | Cause-effect diagram | Not considered | Not available |
| (De Assunçao, Di Costanzo and Buyya 2009) | Evaluating the optimality of scheduling strategies used by an organisation to reduce response time in using IaaS. | Type V | Enable phase | Organisation | Performance metrics | Not specified | Not specified | Not specified | Not considered | Not available |
| (Deelman, Singh, Livny et al. 2008) | Analysing the cost-performance trade-off between difference executions and resource provisioning plans by legacy systems. | Type III | Enable phase | Legacy systems | Performance metrics | Not specified | Not specified | Not specified | Not considered | Not available |
| (Kondo, Javadi, Malecot et al. 2009) | Comparing cost and performance of legacy systems in using IaaS. | Type V | Enable phase | Legacy systems | Performance metrics | Not specified | Not specified | Not specified | Not considered | Not available |
| (Walker, Brisken and Romney 2010) | Reasoning about the cost of leasing infrastructure from cloud storage. | Type III | Enable phase | Organisation | Net present value | Not specified | Not specified | Not specified | Not considered | Not available |
| (Garg, Versteeg and Buyya 2011) | Measuring, comparing, and prioritizing cloud services based on users' requirements. | Type V | Enable phase | Organisation/ Legacy systems | QoS metrics | Yes | Implicitly supported | Not specified | Not considered | Not available |
| (Godse and Mulik 2009) | Analysing and selecting appropriate SaaS products. | Type II | Enable phase | Organisation | AHP | Embedded in framework description | Implicitly supported | Not specified | Not considered | Not available |
| (Menzel, Schönherr and Tai 2013) | Examining if IaaS meets organisation's needs by evaluating and ranking alternatives using a set of criteria catalogue. | All | Enable phase | Organisation/ Legacy systems | ANP | Yes | Implicitly supported | Not specified | Not considered | Yes |
| (Scandurra, Mongiello, Colucci et al. 2016) | Redeploying e-commerce cloud applications on different servers at run-time based on evolving requirements, sudden changes in the operational environment conditions, and application traffic. | | Maintain phase | Legacy systems | Goal reasoning | Embedded in framework description | Not specified | Graph modelling | Not considered | Not available |
| (Nikkhouy 2013) | Exploring potential benefits and risks in migrating legacy systems to cloud services. | All | Plan phase | Organisation | Change analysis | Yes | Implicitly supported | Cause and effect diagram | Not considered | Not available |
| (Christoforou and Andreou 2013) | Assessing the feasibility of the cloud adoption in organizations regarding factors such as security, legal issues, availability, cost, return on investment (ROI), compliance, performance, scalability, and data access/import-export. | All | Plan phase | Organisation | Analysing influencing factors | Embedded in framework description | Yes | Influence diagrams modelling | Not considered | Not available |
| (Tak, Urgaonkar and Sivasubramaniam 2011) | Exploring factors such as workload intensity, growth rate, storage capacity and software licensing costs affecting the cost of deployment options in the cloud. | V | Plan phase | Legacy systems | Using benchmarks representing of different scenarios | Not specified | Not specified | NPV models | Not considered | Not available |
| (Juan-Verdejo and Baars 2013) | Identifying suitable components of legacy systems for deploying in IaaS with respect to interdependencies among components and factors such as data transfer volumes, performance, sensitivity of cloud-based data repositories, and exposure to public networks. | V | Enable phase | Legacy systems | Combination of scenario based & AHP | Embedded in framework description | Yes | Legacy system architecture model | Not considered | Not available |
| (Menzel and Ranjan 2012) | Identifying a compatible combination of software images (e.g., Web server image) in mapping web applications to virtualized cloud services while expected QoS of applications are satisfied. | V | Enable phase | Legacy systems | Combination of optimization and AHP | Embedded in framework description | Yes | Simulation models | Not considered | Not available |
| (Fittkau, Frey and Hasselbring 2012) | Evaluation of competing cloud deployment options and finding the most suitable mapping of virtual machines to cloud services regarding cost and system performance. | V | Enable phase | Legacy systems | Combination of optimization and scenario-based | Yes | Yes | Simulation models | Not considered | Not available |
| (Saripalli and Pingali 2011) | Ranking legacy system workloads for migrating to cloud environments based on attributes such as latency, bandwidth, and cost. | All | Enable phase | Legacy systems | Combination of multi-attribute decision making and wide-band Delphi | Embedded in framework description | Yes | Decision matrix | Not considered | Not available |
| (Calheiros, Ranjan, Beloglazov et al. 2011) | Determining the best deployment options of legacy system components of on cloud servers whilst QoS are satisfied. | V | Enable phase | Legacy systems | Scenario-based | Embedded in framework description | Yes | Simulation models | Not considered | Yes |
| (Leymann, Fehling, Mietzner et al. 2011) | Rearrangement of the legacy application deployment topology in cloud servers regarding dependencies among its components and requirements such as latency, transfer, and data privacy are addressed. | V | Enable phase | Legacy systems | Optimisation algorithm (e.g. simulated annealing) | Yes | Not specified | Metamodeling, application templates | Not considered | Yes |
| (Giovanoli 2012) | Assessing and selecting the most suitable cloud services via guidelines provided in a database containing information of different cloud service providers. | All | Enable phase | Legacy systems | Not specified | Not specified | Not specified | Not specified | Information repository of cloud service providers | Yes |
| (Zardari, Bahsoon and Ekárt 2014) | Prioritising obstacles related to cloud service adoption and resolution tactics. | All | Plan phase | Legacy systems | Goal reasoning and AHP | Yes | Yes | Goal models | Not considered | No |
| This work | Analysing goal-obstacle in migrating legacy systems to cloud platforms along with utilization of an evidence-based repository during the steps of the goal-oriented elaboration process. | All | Plan and enable phases | Legacy systems | Goal reasoning and evidence-based approach | Yes | Yes | Goal models | Yes | No |

## 7. Research contributions

Firstly, the proposed framework enables system architects to make context driven decision on adopting cloud services rather than merely on the basis of their novelty or available anecdotal evidence. The repository component of the framework is, in essence, a knowledge sharing platform. It strives providing a body of documented evidence from the extant literature. This body of knowledge informs cloud adoption requirement analysis ultimately enhancing the reliability and any concomitant decision.

Secondly, an early stage analysis of cloud migration goals is not trivial. There is a dearth of research on how to elicit, model, and anticipate potential impacts of obstacles on them in a systematic way. We provided a systematic framework to explore goals, exceptional conditions impeding these goals, and to produce a complete set of requirements. The framework has been built on top of the empirical knowledge that makes results of goal-obstacle analysis more reliable compared to a situation in which the analysis is merely based on general knowledge of cloud platforms or personal experience of the system architect. The output from the framework is a goal-oriented requirements model relating cloud migration goals to risky obstacles following with operational countermeasures. This model gives the system architects a broad view of rationale and costs of specific requirements before delving into technical aspects of integrating systems with cloud services. The model can be incorporated into the implementation stage to make appropriate trade-offs on the basis of, for example, cost, security, or performance goals. Not only the framework applicability is positioned in the earliest stage of migration, but it can also be used during the post-migration stage to tackle costly mistakes.

Finally, the framework can be employed to complement existing decision making frameworks as reviewed in Section 6. It fills their gaps in reusing existing empirical knowledge and the strategic goals of the overall migration process. It can be also used as a stand-alone framework for a goal-obstacle analysis of cloud migration types to reason about risky obstacles. Our framework takes a qualitative approach as its aim is not to quantitatively measure probability occurrence of obstacles or goal achievement, instead; the framework simply intents in specifying cloud migration goals that might be impacted by obstacles. We hope the current study provides a motivation for combining the evidence-based software engineering and goal-oriented modelling literature and stimulates more efforts in the context of cloud computing.

## 8. Threats to validity

Our framework has been validated to account for both internal and external validity threats. *Internal validity threats* relate to factors that a researcher has not been aware of and may have affected the research outcome i.e. the framework artefact itself (Wohlin, Runeson, Höst et al. 2012). *External validity* threats relate to the extent to which the resulting framework can be generalised (Wohlin, Runeson, Höst et al. 2012).

To ensure the repository's coverage, we focused on studies pertaining to legacy systems transition to cloud platforms in the SLR depicted in Figure 2. The SLR identified 112 studies and two itemized collections, respectively, presented in Appendix A and B. SLRs are generally criticised for being too mechanical, protocol-driven, and formal that limits research's curiosity and scholarly examining of knowledge in a literature review (Hjørland 2011; Boell and Cecez-Kecmanovic 2015). Another common concern associated with SLRs is their indeterminacy and multiplicity of a domain language. This latter concern is of particular relevance to the cloud computing field where precise terminologies or nomenclature have not yet been grounded. A particular obstacle or resolution tactic may be expressed using different terms and vocabularies. For example, we found that studies in Appendix A do not necessarily use the search strings presented in Table 1 or terms goals, obstacles, resolution tactics, and decision making. To mitigate against the incompleteness of the framework repository, e.g. due to missing some important and reusable empirical findings, our SLR had a phase for early understanding and critical reading of the cloud migration literature before it is fine-tuned as shown in Figure 2. For example, we did not confine ourselves with the fixed search strings presented in Table 1; rather, we sought concepts related to goals, obstacles, resolution tactics and not merely for search strings because such concepts were not only expressed using search strings and sometimes they were described or paraphrased. With all due care taken above in conducting the SLR, it is not possible to affirm that the repository is complete.

Additionally, the reliability of case studies is subjected to the quality and accuracy of the written documents of them. The documents used for the case studies may have been slightly different from activities that actually had been performed due to reasons such as hindsight bias or error in remembering details. As a consequence, there is a possibility of missing the identification of some new obstacles and resolution tactics that could be added as new entry to the framework repository or change the procedure's step of the framework. This may have weakened internal validity of running case studies. To mitigate against this, we conducted follow-up communications with key document providers to confirm the validity of the documents of projects and to provide any missing information.

Furthermore, an often-cited limitation of case studies is their specificity to a particular context at a particular point in time which circumscribes generalisability of results to other applications and contexts. Although the framework was validated through two idiosyncratic case studies, its applicability to all possible cases can still, of course, be debated. The repository is however extensible with new entries if more case studies are performed.

Finally, we do not claim the framework procedure is complete to provide a great analysis of all scenarios of transition cloud platforms. There might be some short-cuts to satisfy goals, or some hidden factors that hinder certain goal achievement but are not detected in the framework procedure. At this stage, there is no assertion regarding the generalisability of the procedure beyond the cases investigated in this study. But it can be extended with new steps if the framework is appraised with more case studies in a variety of scenarios.

## 9. Conclusion and future work

This article is based on this premise that endeavours towards cloud migration are sometimes rewarding or challenging along with many lessons learned along the way. Reusing these lessons in different scenarios is a promising approach in a better exploration of uncertain risks against cloud adoption goals and reliability of decision outcomes. In this regard, our proposed framework harnesses a synergy between evidence-based software engineering and goal-oriented modelling approaches. The proposed framework comprises an itemized evidence-based repository and a cloud adoption goal-obstacle procedure utilizing the repository information. This is the first attempt turning the existing body of knowledge of cloud enablement into a concise, accessible, and a reusable source. This has not been a feature of the past research. Nevertheless, some deficiencies regarding the completeness of the repository are clear areas for further research as discussed in the following.

Firstly, there is an unequal availability of empirical studies in the literature in support of the repository collections. On the one hand, as shown in the Appendix B and suggested by several studies, the resolution tactic *Develop adaptor/wrapper (T6)* can be used in addressing several obstacles namely *Incompatible pluggable cloud services (O19)*, *Incomplete APIs (O20)*, *Incompatible data types (O21)*, *Operating system incompatibility (O22)*, *Machine-image incompatibility (O23)*, *Virtual machine contextualization incompatibility (O24)*, *API incompatibility across multiple cloud (O25)*, and *Proprietary APIs (O36)*. On the other hand, there is only one resolution tactic to address the obstacle *Extra testing effort (O32)* which is *Prioritize tests (T30)*. Hence, further research is required to add more empirical findings to the repository as more studies appear in the cloud computing literature.

Secondly, we plan to add a probabilistic layer for goal specification and obstacle assessment in view of their estimation and required degrees of satisfaction grounded on system domain. The criticality of obstacle consequences will be computed by propagation probabilities from leaf obstacles towards high-level goals through the goal refinement model. To this aim, we will extend the procedure's steps of the framework by annotating obstacle and goal elements with the probability of their occurrence (Cailliau and van Lamsweerde 2013).

Finally, the framework repository in its current state is stored in the textual template and does not provide a systematic mechanism for regularly updating the repository with new empirical data as identified in the literature. Also, the goal-obstacle procedure utilizing the repository is manual. These deficiencies confine the usability of the framework. We plan to provide a tool support that facilitates using the framework when working with large-scale goal models.

# Reference

A. Aurum, R. J., C. Wohlin, and M. Handzic (2003). Managing Software Engineering Knowledge.

AmazonS3 "Amazon Web Services S3 - Simple Cloud Storage Service." https://aws.amazon.com/s3/?sc_channel=PS&sc_campaign=acquisition_AU&sc_publisher=google&sc_medium=s3_b&sc_content=s3_e&sc_detail=s3.amazonaws.com&sc_category=s3&sc_segment=118649900484&sc_matchtype=e&sc_country=AU&s_kwcid=AL!4422!3!118649900484!e!!g!!s3.amazonaws.com&ef_id=Uvy8OgAABEYZb5Xa:20161022040726:s (Last accessed October 2016).

Anstett, T., F. Leymann, R. Mietzner, et al. (2009). Towards BPEL in the Cloud: Exploiting Different Delivery Models for the Execution of Business Processes. Services - I, 2009 World Conference on.

Babar, M. A., Z. Liming and R. Jeffery (2004). A framework for classifying and comparing software architecture evaluation methods. Software Engineering Conference, 2004. Proceedings. 2004 Australian.

Boell, K. S. and D. Cecez-Kecmanovic (2015). "On being 'systematic' in literature reviews in IS." Journal of Information Technology **30**(2): 161-173.

Cailliau, A. and A. van Lamsweerde (2013). "Assessing requirements-related risks through probabilistic goals and obstacles." Requirements Engineering **18**(2): 129-146.

Calheiros, R. N., R. Ranjan, A. Beloglazov, et al. (2011). "CloudSim: a toolkit for modeling and simulation of cloud computing environments and evaluation of resource provisioning algorithms." Software: Practice and Experience **41**(1): 23-50.

Chow, R., P. Golle, M. Jakobsson, et al. (2009). Controlling data in the cloud: outsourcing computation without outsourcing control. Proceedings of the 2009 ACM workshop on Cloud computing security, ACM.

Christoforou, A. and A. S. Andreou (2013). A Cloud Adoption Decision Support Model Using Influence Diagrams. Artificial Intelligence Applications and Innovations, Springer: 151-160.

Dardenne, A., A. Van Lamsweerde and S. Fickas (1993). "Goal-directed requirements acquisition." Science of computer programming **20**(1): 3-50.

De Assunçao, M. D., A. Di Costanzo and R. Buyya (2009). Evaluating the cost-benefit of using cloud computing to extend the capacity of clusters. Proceedings of the 18th ACM international symposium on High performance distributed computing, ACM.

Deelman, E., G. Singh, M. Livny, et al. (2008). The cost of doing science on the cloud: the montage example. Proceedings of the 2008 ACM/IEEE conference on Supercomputing, IEEE Press.

Dieste, O. and O. Padua (2007). Developing search strategies for detecting relevant experiments for systematic reviews. Empirical Software Engineering and Measurement, 2007. ESEM 2007. First International Symposium on, IEEE.

Dragoni, N., S. Giallorenzo, A. L. Lafuente, et al. (2016). "Microservices: yesterday, today, and tomorrow." arXiv preprint arXiv:1606.04036.

Dyba, T., B. A. Kitchenham and M. Jorgensen (2005). "Evidence-based software engineering for practitioners." Software, IEEE **22**(1): 58-65.

El-Gazzar, R., E. Hustad and D. H. Olsen (2016). "Understanding cloud computing adoption issues: A Delphi study approach." Journal of Systems and Software **118**: 64-84.

Fahmideh, M., F. Daneshgar, G. Beydoun, et al. (2017). "Challenges in migrating legacy software systems to the cloud — an empirical study." Information Systems **67**(Supplement C): 100-113.

Fahmideh, M., F. Daneshgar, G. Low, et al. (2016). "Cloud migration process—A survey, evaluation framework, and open challenges." Journal of Systems and Software **120**: 31-69.

Fahmideh, M., Low Graham, Ghassan Beydoun (2016b). "Conceptualising Cloud Migration Process." Twenty-Fourth European Conference on Information Systems (ECIS), İstanbul,Turkey, 2016 **1**: 0.

Fittkau, F., S. Frey and W. Hasselbring (2012). CDOSim: Simulating cloud deployment options for software migration support. Maintenance and Evolution of Service-Oriented and Cloud-Based Systems (MESOCA), 2012 IEEE 6th International Workshop on the.

Franklin, C. (1996). "Lt. Gen (USAF) Commander ESC, January 1996, Memorandum for ESC Program Managers, ESC/CC." Risk Management, Department of the Air Force, Headquarters ESC (AFMC) Hanscom Air Force Base, MA.

Garg, S. K., S. Versteeg and R. Buyya (2011). SMICloud: A Framework for Comparing and Ranking Cloud Services. Utility and Cloud Computing (UCC), 2011 Fourth IEEE International Conference on.

Garg, S. K., S. Versteeg and R. Buyya (2013). "A framework for ranking of cloud computing services." Future Generation Computer Systems **29**(4): 1012-1023.

Giovanoli, G. (2012). Building a Knowledge Base for Guiding Users through the Cloud Life Cycle

Godse, M. and S. Mulik (2009). An approach for selecting software-as-a-service (SaaS) product. Cloud Computing, 2009. CLOUD'09. IEEE International Conference on, IEEE.

Godse, M. and S. Mulik (2009). An approach for selecting software-as-a-service (SaaS) product. 2009 IEEE International Conference on Cloud Computing, IEEE.

Gordon, J. (2015). Case Study: Refactoring A Monolith Into A Cloud Native App, https://www.dropbox.com/s/t0sfl6cn6uounhy/Migration%20scenario-SpringTrader.docx?dl=0.

Gregor, S. and D. Jones (2007). "The anatomy of a design theory." Journal of the Association for Information Systems **8**(5): 312-335.

Henver, A., S. T. March, J. Park, et al. (2004). "Design science in information systems research." MIS quarterly **28**(1): 75-105.

Hjørland, B. (2011). "Evidence-based practice: An analysis based on the philosophy of science." Journal of the American society for information science and technology **62**(7): 1301-1310.

Juan-Verdejo, A. and H. Baars (2013). Decision support for partially moving applications to the cloud: the example of business intelligence. Proceedings of the 2013 international workshop on Hot topics in cloud services. Prague, Czech Republic, ACM**:** 35-42.

Khajeh-Hosseini, A., D. Greenwood and I. Sommerville (2010). Cloud Migration: A Case Study of Migrating an Enterprise IT System to IaaS. Cloud Computing (CLOUD), 2010 IEEE 3rd International Conference on.

Khajeh-Hosseini, A., I. Sommerville, J. Bogaerts, et al. (2011). Decision support tools for cloud migration in the enterprise. Cloud Computing (CLOUD), 2011 IEEE International Conference on, IEEE.

Khajeh-Hosseini, A., D. Greenwood, J. W. Smith, et al. (2012). "The cloud adoption toolkit: supporting cloud adoption decisions in the enterprise." Software: Practice and Experience **42**(4): 447-465.

Kitchenham, B., O. Pearl Brereton, D. Budgen, et al. (2009). "Systematic literature reviews in software engineering – A systematic literature review." Information and software technology **51**(1): 7-15.

Kondo, D., B. Javadi, P. Malecot, et al. (2009). Cost-benefit analysis of cloud computing versus desktop grids. Parallel & Distributed Processing, 2009. IPDPS 2009. IEEE International Symposium on, IEEE.

Lacity, M. C., S. A. Khan and L. P. Willcocks (2009). "A review of the IT outsourcing literature: Insights for practice." The Journal of Strategic Information Systems **18**(3): 130-146.

Letier, E. (2001). Reasoning about agents in goal-oriented requirements engineering, PhD thesis, Université catholique de Louvain.

Leymann, F., C. Fehling, R. Mietzner, et al. (2011). "Moving applications to the cloud: An approach based on application model enrichment." International Journal of Cooperative Information Systems **20**(03): 307-356.

Linthicum, D. (2012). "Why Cloud Computing Projects Fail?" Available at: http://www.slideshare.net/Linthicum/why-cloud-computing-projects-fail, last access October 2016.

Low, C., Y. Chen and M. Wu (2011). "Understanding the determinants of cloud computing adoption." Industrial management & data systems **111**(7): 1006-1023.

Menzel, M. and R. Ranjan (2012). CloudGenius: decision support for web server cloud migration. Proceedings of the 21st international conference on World Wide Web, ACM.

Menzel, M., M. Schönherr and S. Tai (2013). "(MC2)2: criteria, requirements and a software prototype for Cloud infrastructure decisions." Software: Practice and Experience **43**(11): 1283-1297.

Nikkhouy, E. (2013). "Decision Making About Migrating To The Cloud Model." Cloud-Based Software Engineering: 8.

Patton, M. Q. (1990). Qualitative evaluation and research methods, SAGE Publications, inc.

Peffers, K., T. Tuunanen, M. A. Rothenberger, et al. (2008). "A design science research methodology for information systems research." Journal of management information systems **24**(3): 45-77.

Pepitone, J. (2011). "Amazon EC2 outage downs Reddit, Quora." Retrieved May **17**: 2011.

Rabetski, P. (2012). "Migration of an on-premise application to the cloud."

Rabetski, P. and G. Schneider (2013). Migration of an On-Premise Application to the Cloud: Experience Report. Service-Oriented and Cloud Computing, Springer: 227-241.

Saripalli, P. and G. Pingali (2011). MADMAC: Multiple Attribute Decision Methodology for Adoption of Clouds. Cloud Computing (CLOUD), 2011 IEEE International Conference on.

Scandurra, P., M. Mongiello, S. Colucci, et al. (2016). Towards a goal-oriented approach to adaptable re-deployment of cloud-based applications. Proceedings of the 6th international conference on cloud computing and services science.

Tak, B. C., B. Urgaonkar and A. Sivasubramaniam (2011). To move or not to move: the economics of cloud computing. Proceedings of the 3rd USENIX conference on Hot topics in cloud computing. Portland, OR, USENIX Association: 5-5.

Thönes, J. (2015). "Microservices." IEEE software **32**(1): 116-116.

Tsidulko, J. (2016). "The 10 Biggest Cloud Outages Of 2016." Available at http://www.crn.com/slide-shows/cloud/300081477/the-10-biggest-cloud-outages-of-2016-so-far.htm.

Van Lamsweerde, A. (2009). "Requirements engineering: from system goals to UML models to software specifications."

van Lamsweerde, A. and E. Letier (2000). "Handling obstacles in goal-oriented requirements engineering." Software Engineering, IEEE Transactions on **26**(10): 978-1005.

Van Lamsweerde, A. and E. Letier (2004). From object orientation to goal orientation: A paradigm shift for requirements engineering. Radical Innovations of Software and Systems Engineering in the Future, Springer: 325-340.

Walker, E., W. Brisken and J. Romney (2010). "To lease or not to lease from storage clouds." Computer **43**(4): 44-50.

Wohlin, C., P. Runeson, M. Höst, et al. (2012). Experimentation in software engineering, Springer Science & Business Media.

Wu, W.-W. (2011). "Mining significant factors affecting the adoption of SaaS using the rough set approach." Journal of Systems and Software **84**(3): 435-441.

Wu, W.-W., L. W. Lan and Y.-T. Lee (2011). "Exploring decisive factors affecting an organization's SaaS adoption: A case study." International Journal of Information Management **31**(6): 556-563.

Yang, H. and M. Tate (2012). "A descriptive literature review and classification of cloud computing research." Communications of the Association for Information systems **31**(2): 35-60.

Yu, E. S. K. (1997). Towards modelling and reasoning support for early-phase requirements engineering. Requirements Engineering, 1997., Proceedings of the Third IEEE International Symposium on.

Zardari, S., R. Bahsoon and A. Ekárt (2014). "Cloud Adoption: Prioritizing Obstacles and Obstacles Resolution Tactics Using AHP."

Zimmermann, O., L. Wegmann, H. Koziolek, et al. (2015). Architectural decision guidance across projects-problem space modeling, decision backlog management and cloud computing knowledge. Software Architecture (WICSA), 2015 12th Working IEEE/IFIP Conference on, IEEE.

## Appendix A (Studies used to develop the knowledge repository of the framework)

| Identifier | Study | Year |
|---|---|---|
| [S1] | Torbacki, W, *SaaS–direction of technology development in ERP/MRP systems*. Archives of Materials Science 58: 58. | 2008 |
| [S2] | Fox, A., *Above the clouds: A Berkeley view of cloud computing*. Dept. Electrical Eng. and Comput. Sciences, University of California, Berkeley,Rep. UCB/EECS 28. | 2009 |
| [S3] | Habib, S. M., *Cloud computing landscape and research challenges regarding trust and reputation*. Ubiquitous Intelligence & Computing and 7th International Conference on Autonomic & Trusted Computing (UIC/ATC), 2010 7th International Conference on, IEEE. | 2010 |
| [S4] | Wood, T, *Disaster recovery as a cloud service: Economic benefits & deployment challenges*. 2nd USENIX Workshop on Hot Topics in Cloud Computing. | 2010 |
| [S5] | Marston, S., *Cloud computing—The business perspective*. Decision support systems 51(1): 176-189. | 2011 |
| [S6] | Anstett, T., *Towards BPEL in the Cloud: Exploiting Different Delivery Models for the Execution of Business Processes*. Services - I, 2009 World Conference on. | 2009 |
| [S7] | Herbert, L., *The ROI Of Software-As-A-Service*. Forrester Research. | 2009 |
| [S8] | Wada, H., *Data Consistency Properties and the Trade-offs in Commercial Cloud Storage: the Consumers' Perspective*, CIDR. | 2011 |
| [S9] | La, H. J., *Technical challenges and solution space for developing SaaS and mash-up cloud services*. e-Business Engineering, ICEBE'09. IEEE International Conference on, IEEE. | 2009 |
| [S10] | Duipmans, E., *Business Process Management in the cloud: Business Process as a Service (BPaaS)*. University of Twente. | 2012 |
| [S11] | Widera P, K. N. , *Protein models comparator: scalable bioinformatics computing on the Google App Engine platform*. Computing Research Repository: 8. | 2011 |
| [S12] | Andrikopoulos, V., *How to adapt applications for the Cloud environment*. Computing 95(6): 493-535. | 2013 |
| [S13] | Brebner, P. C., *Is your cloud elastic enough?: performance modelling the elasticity of infrastructure as a service (IaaS) cloud applications*. Proceedings of the third joint WOSP/SIPEW international conference on Performance Engineering, ACM. | 2012 |
| [S14] | Li, A., *CloudCmp: comparing public cloud providers*. Proceedings of the 10th ACM SIGCOMM conference on Internet measurement, ACM. | 2010 |
| [S15] | Mansfield-Devine, S, *Cloud Security: Danger in the clouds*. Netw. Secur. 2008(12): 9-11. | 2008 |
| [S16] | Jensen, M., *On technical security issues in cloud computing*. Cloud Computing, 2009. CLOUD'09. IEEE International Conference on, IEEE. | 2009 |
| [S17] | Hay, B., *Storm clouds rising: security challenges for IaaS cloud computing*. System Sciences (HICSS), 2011 44th Hawaii International Conference on, IEEE. | 2011 |
| [S18] | Vogels, W, *Eventually consistent*. Communications of the ACM 52(1): 40-44. | 2009 |
| [S19] | Ristenpart, T., *Hey, you, get off of my cloud: exploring information leakage in third-party compute clouds*. Proceedings of the 16th ACM conference on Computer and communications security, ACM. | 2009 |
| [S20] | Gupta, R., *Above the Clouds: A View of Cloud Computing*. Asian Journal of Research in Social Sciences and Humanities 2(6): 84-110. | 2012 |
| [S21] | Zissis, D, *Addressing cloud computing security issues*. Future Generation Computer Systems 28(3): 583-592. | 2012 |
| [S22] | Hubbard, D., *Top Threats to Cloud Computing V1. 0*. Cloud Security Alliance. | 2010 |
| [S23] | Hussain, O. K., *A framework for user feedback based cloud service monitoring*. Complex, Intelligent and Software Intensive Systems (CISIS), Sixth International Conference on, IEEE. | 2012 |
| [S24] | Alhamad, M., *Conceptual SLA framework for cloud computing*. Digital Ecosystems and Technologies (DEST), 4th IEEE International Conference on, IEEE. | 2010 |
| [S25] | Harmer, T., *Provider-Independent Use of the Cloud*. Euro-Par 2009 Parallel Processing. H. Sips, D. Epema etal., Springer Berlin Heidelberg. 5704: 454-465. | 2009 |
| [S26] | K. Keahey, Matsunaga, and J. Fortes. *Sky Computing*. IEEE Internet Computing, Palo Alto vol. 13: 315-340. | 2009 |
| [S27] | Loutas, N., *Towards a Reference Architecture for Semantically Interoperable Clouds*. Cloud Computing Technology and Science (CloudCom), IEEE Second International Conference on. | 2010 |
| [S28] | Martinez Garro, *Constructing hybrid architectures and dynamic services in Cloud BPM*. Science and Information Conference (SAI). | 2013 |
| [S29] | Zissis, D., *Addressing cloud computing security issues*. Future Generation Computer Systems 28(3): 583-592. | 2012 |
| [S30] | Vecchiola, C, *Deadline-driven provisioning of resources for scientific applications in hybrid clouds with Aneka*. Future Generation Computer Systems 28(1): 58-65. | 2012 |
| [S31] | Kossmann, D, *Data Management in the Cloud: Promises, State-of-the-art, and Open Questions*. Datenbank-Spektrum 10(3): 121-129. | 2010 |
| [S32] | Iosup, A., *On the performance variability of production cloud services*. Cluster, Cloud and Grid Computing (CCGrid), 11th IEEE/ACM International Symposium on, IEEE. | 2011 |
| [S33] | Ma, D. *The Business Model of Software-As-A-Service*. Services Computing, 2007. SCC 2007. IEEE International Conference on, IEEE. | 2007 |
| [S34] | Buyya, R., *Market-oriented cloud computing: Vision, hype, and reality for delivering it services as computing utilities*. High Performance Computing and Communications, HPCC'08. 10th IEEE International Conference on, Ieee. | 2008 |
| [S35] | Rayport, J. F, *Envisioning the cloud: the next computing paradigm*. Int. J. Database Manage. Syst.(IJDMS) 1(1). | 2009 |
| [S36] | Vogels, W., *CTO roundtable: cloud computing*. | 2009 |
| [S37] | Gao, J., *Cloud testing-issues, challenges, needs and practice*. Software Engineering: An International Journal 1(1): 9-23. | 2011 |
| [S38] | Strauch, S., *Using Patterns to Move the Application Data Layer to the Cloud*. Proceedings of the 5th International Conference on Pervasive Patterns and Applications, PATTERNS 2013, 27 May – June 1 2013, Valencia, Spain, Xpert Publishing Services (XPS). | 2013 |
| [S39] | Armstrong, D., *Towards a contextualization solution for cloud platform services*. Cloud Computing Technology and Science (CloudCom), 2011 IEEE Third International Conference on, IEEE. | 2011 |
| [S40] | Sriram, I. *Research agenda in cloud technologies*. arXiv preprint arXiv:1001.3259. | 2010 |
| [S41] | Batarseh, F. A, *Context-assisted test cases reduction for cloud validation*. International and Interdisciplinary Conference on Modeling and Using Context, Springer. | 2013 |
| [S42] | Parveen, T., *When to Migrate Software Testing to the Cloud?* Software Testing, Verification, and Validation Workshops (ICSTW), 2010 Third International Conference on. | 2010 |
| [S43] | Tran, V., *Application migration to cloud: a taxonomy of critical factors*. Proc. of 2nd International Workshop on Software Engineering for Cloud Computing, ACM. | 2011 |
| [S44] | Khajeh-Hosseini, A, *Cloud migration: A case study of migrating an enterprise it system to iaas*. Cloud Computing (CLOUD), IEEE 3rd International Conference on, IEEE. | 2010 |
| [S45] | Catteddu, D. (2010). *Cloud Computing: Benefits, Risks and Recommendations for Information Security*. Web Application Security. C. Serrão, V. Aguilera Díaz and F. Cerullo, Springer Berlin Heidelberg. 72: 17-17. | 2010 |
| [S46] | Khajeh-Hosseini, A., *Research challenges for enterprise cloud computing*. arXiv preprint arXiv:1001.3257. | 2010 |
| [S47] | Torbacki, W. (2008). *SaaS–direction of technology development in ERP/MRP systems*. Archives of Materials Science 58: 58. | 2008 |
| [S48] | M. Xin, N. L. *Software-as-a-service model: elaborating client-side adoption factors*. Proceedings of the Twenty-ninth International Conference on Information | 2008 |
| [S49] | Kerr, J., *Cloud computing: legal and privacy issues*, Journal of Legal Issues and Cases in Business 1: 1. | 2012 |
| [S50] | Leavitt, N, *Is Cloud Computing Really Ready for Prime Time?* Computer 42(1): 15-20. | 2009 |
| [S51] | Satzger, B., et al. (2013). *Winds of change: from vendor lock-in to the meta cloud*. Internet Computing, IEEE 17(1): 69-73. | 2012 |
| [S52] | Silva, G. C, *A systematic review of cloud lock-in solutions*, Cloud Computing Technology and Science (CloudCom), 2013 5th International Conference on, IEEE. | 2013 |
| [S53] | Dillon, T., *Cloud computing: Issues and challenges*. Advanced Information Networking and Applications (AINA), 24th IEEE International Conference on, Ieee. | 2010 |
| [S54] | So, K, *Cloud computing security issues and challenges*. International Journal of Computer Networks 3(5). | 2011 |
| [S55] | Dalheimer, M., *Genlm: license management for grid and cloud computing environments*. Cluster Computing and the Grid, CCGRID'09. 9th IEEE/ACM International Symposium on, IEEE. | 2009 |
| [S56] | Morgan, L., *Factors affecting the adoption of cloud computing: an exploratory study*. | 2013 |
| [S57] | Joint, A., *Hey, you, get off of that cloud?* Computer Law & Security Review 25(3): 270-274. | 2009 |
| [S58] | Hajjat, M., *Cloudward bound: planning for beneficial migration of enterprise applications to the cloud*. Proceedings of the ACM SIGCOMM conference. New Delhi, India, ACM: 243-254. | 2010 |
| [S59] | Bezemer, C.-P. *Multi-tenant SaaS applications: maintenance dream or nightmare?* Proceedings of the Joint ERCIM Workshop on Software Evolution (EVOL) and International Workshop on Principles of Software Evolution (IWPSE), ACM. | 2010 |
| [S60] | Chong, F., *Multi-tenant data architecture*. MSDN Library, Microsoft Corporation | 2006 |
| [S61] | Krebs, R., *Architectural Concerns in Multi-tenant SaaS Applications*. CLOSER. | 2012 |
| [S62] | Mietzner, R., *Variability modeling to support customization and deployment of multi-tenant-aware Software as a Service applications*. Proceedings of the 2009 ICSE Workshop on Principles of Engineering Service Oriented Systems, IEEE Computer Society: 18-25. | 2009 |
| [S63] | Gonidis, F., *Addressing the challenge of application portability in cloud platforms*. 7th South-East European Doctoral Student Conference. | 2012 |
| [S64] | Petcu, D. *Portability and interoperability between clouds: challenges and case study*. European Conference on a Service-Based Internet, Springer. | 2011 |
| [S65] | Chauhan, M. A., *Towards Process Support for Migrating Applications to Cloud Computing*. Cloud and Service Computing (CSC), International Conference on. | 2012 |
| [S66] | Strauch, S., *Migrating eScience Applications to the Cloud: Methodology and Evaluation*. | 2014 |
| [S67] | Strauch, S., *Migrating Enterprise Applications to the Cloud: Methodology and Evaluation*, International Journal of Big Data Intelligence. | 2014 |
| [S68] | ShaoJie, T., AMAZING: An Optimal Bidding Strategy for Amazon EC2 Cloud Spot Instance, available at http://www.cs.iit.edu/~xli/paper/Conf/EC-CLOUD2012.pdf | 2016 |
| [S69] | Computing, C., *Toward a multi-tenancy authorization system for cloud services*. | 2010 |
| [S70] | Karampaglis, Z., *Secure Migration of Legacy Applications to the Web*. Migration(1/18). | 2012 |
| [S71] | Pahl, C., *A Comparison of On-Premise to Cloud Migration Approaches*. Service-Oriented and Cloud Computing, Springer: 212-226. | 2013 |
| [S72] | Council, C. S. C. , *Migration applications to public Cloud Services: roadmap for success*. | 2013 |
| [S73] | Menychtas, A., *ARTIST Methodology and Framework: A novel approach for the migration of legacy software on the Cloud*. Symbolic and Numeric Algorithms for Scientific Computing (SYNASC), 15th International Symposium on, IEEE. | 2013 |
| [S74] | Varia, J., *Migrating your existing applications to the aws cloud*: *A Phase-driven Approach to Cloud Migration*. | 2010 |
| [S75] | Betts, D., *Moving Apps to the Cloud on Microsoft* | 2012 |
| [S76] | Binz, T., *CMotion: A framework for migration of applications into and between clouds*. Service-Oriented Computing and Applications (SOCA), 2011 IEEE International Conference on. | 2011 |
| [S77] | Rabetski, P., *Migration of an On-Premise Application to the Cloud: Experience Report*. Service-Oriented and Cloud Computing, Springer: 227-241. | 2013 |
| [S78] | Bahga, A., *Rapid Prototyping of Multitier Cloud-Based Services and Systems*. Computer 46(11): 76-83. | 2013 |
| [S79] | Chauhan, M. A., *Migrating Service-Oriented System to Cloud Computing: An Experience Report*. Cloud Computing (CLOUD), IEEE International Conference on. | 2011 |
| [S80] | Guo, C. J., *A framework for native multi-tenancy application development and management*. E-Commerce Technology and the 4th IEEE International Conference on Enterprise Computing, E-Commerce, and E-Services, 2007. CEC/EEE 2007. The 9th IEEE International Conference on, IEEE. | 2007 |
| [S81] | Huru, H. A., *MILAS: ModernIzing Legtacy Applications towards Service Oriented Architecture (SOA) and Software as a Service (SaaS)*. | 2009 |
| [S82] | Zagarese, Q., *Enabling advanced loading strategies for data intensive web services*. Web Services (ICWS), IEEE 19th International Conference on, IEEE. | 2012 |
| [S83] | Miranda, J., *Assisting Cloud Service Migration Using Software Adaptation Techniques*. Proceedings of the 2013 IEEE Sixth International Conference on Cloud Computing, IEEE Computer Society. | 2013 |
| [S84] | Laszewski, T., *Migrating to the Cloud: Oracle Client/Server Modernization*, Elsevier. | 2011 |
| [S85] | Bessani, A., *DepSky: dependable and secure storage in a cloud-of-clouds*. Proceedings of the sixth conference on Computer systems. Austria, ACM: 31-46. | 2011 |
| [S86] | Nussbaumer, N., *Cloud Migration for SMEs in a Service Oriented Approach. Computer Software and Applications*, Conference Workshops (COMPSACW), IEEE 37th Annual. | 2013 |
| [S87] | Banerjee, J., *Moving to the cloud: Workload migration techniques and approaches*. High Performance Computing (HiPC), 19th International Conference on, IEEE. | 2012 |
| [S88] | Bezemer, C.-P., *Enabling multi-tenancy: An industrial experience report*. Software Maintenance (ICSM), IEEE International Conference on, IEEE. | 2010 |
| [S89] | Kwok, T., *A software as a service with multi-tenancy support for an electronic contract management application*. Services Computing, 2008. SCC'08. IEEE International Conference on, IEEE. | 2008 |
| [S90] | Zhu, Y., *A Platform for Changing Legacy Application to Multi-tenant Model*. International Journal of Multimedia and Ubiquitous Engineering 9(8): 407-418. | 2014 |
| [S91] | Pahl, C., *Migration to PaaS clouds-Migration process and architectural concerns*. Maintenance and Evolution of Service-Oriented and Cloud-Based Systems (MESOCA), 2013 IEEE 7th International Symposium on the, IEEE. | 2013 |
| [S92] | Durkee, D., *Why cloud computing will never be free*. Queue 8(4): 20. | 2010 |
| [S93] | Barker, S. K., *Empirical evaluation of latency-sensitive application performance in the cloud*. Proceedings of the first annual ACM SIGMM conference on Multimedia systems, ACM. | 2010 |
| [S94] | Khajeh-Hosseini, A., *Decision support tools for cloud migration in the enterprise*. Cloud Computing (CLOUD), IEEE International Conference on, IEEE. | 2011 |
| [S95] | Batarseh, F. A., *Context-assisted test cases reduction for cloud validation*. International and Interdisciplinary Conference on Modeling and Using Context, Springer. | 2013 |
| [S96] | Krebs, R., *Architectural Concerns in Multi-tenant SaaS Applications*. CLOSER 12: 426-431. | 2012 |
| [S97] | Strauch, S., *ESB MT: Enabling Multi-Tenancy in Enterprise Service Buses*. Cloud Computing Technology and Science (CloudCom), 2012 IEEE 4th International Conference on, IEEE. | 2012 |
| [S98] | S.Strauch, V. A., *Decision Support for the Migration of the Application Database Layer to the Cloud*. Proceedings of the 5th IEEE International Conference on Cloud Computing Technology and Science, CloudCom 2013, 2-5 December 2013, Bristol, UK: 639--646. | 2013 |
| [S99] | Qaisi, L.M., *A Twitter Sentiment Analysis for Cloud Providers: A Case Study of Azure Vs. Aws*, Computer Science and Information Technology (CSIT), 2016 7th International Conference on: | 2016 |
| | IEEE, pp. 1-6. | |
| [S100] | Dignan, L., *Public Cloud Computing Vendors: A Look at Strengths, Weaknesses, Big Picture*, available at http://www.zdnet.com/article/public-cloud-computing-vendors-a-look-at-strengths-weaknesses-big-picture/. | 2016 |
| [S101] | Tsai, P., *Aws Vs. Azure: It Pros Weigh the Pros and Cons*, available at: https://www.cloudcomputing-news.net/news/2016/sep/06/aws-vs-azure-it-pros-weigh-pros-and-cons. | 2016 |
| [S102] | Serrano, N., *Infrastructure as a Service and Cloud Technologies*, IEEE software (32:2), pp. 30-36. | 2015 |
| [S103] | Modi, C., *A Survey on Security Issues and Solutions at Different Layers of Cloud Computing*, The journal of supercomputing), pp. 1-32. | 2013 |
| [S104] | Somorovsky, J., *All Your Clouds Are Belong to Us: Security Analysis of Cloud Management Interfaces*, Proceedings of the 3rd ACM workshop on Cloud computing security workshop: ACM, pp. 3-14. | 2011 |
| [S105] | Tajadod, G., *Microsoft and Amazon: A Comparison of Approaches to Cloud Security*, Cloud Computing Technology and Science (CloudCom), 2012 IEEE 4th International Conference on: IEEE, pp. 539-544. | 2012 |
| [S106] | Roloff, Eduardo, *Evaluating high performance computing on the windows azure platform*, Cloud Computing (CLOUD), IEEE 5th International Conference on. IEEE. | 2012 |
| [S107] | Google documentation, *Regions and Zones*, available at https://cloud.google.com/compute/docs/regions-zones/regions-zones. | 2017 |
| [S108] | Smeds, J., Nybom, *Devops: A Definition and Perceived Adoption Impediments*, in Agile Processes in Software Engineering and Extreme Programming: 16th International Conference, Xp 2015, Helsinki, Finland, May 25-29, 2015, Proceedings, C. Lassenius, T. Dingsøyr and M. Paasivaara (eds.). Cham: Springer International Publishing, pp. 166-177. | 2015 |
| [S109] | Riungu-Kalliosaari, L., *DevOps Adoption Benefits and Challenges in Practice: A Case Study*, Product-Focused Software Process Improvement: 17th International Conference, PROFES 2016, Trondheim, Norway, November 22-24, 2016, Proceedings 17. Springer International Publishing. | 2016 |
| [S110] | Pahl, Claus, *Containerization and the paas cloud*, IEEE Cloud Computing 2.3, p 24-31. | 2015 |
| [S111] | Gunka, A, *Moving an application to the cloud: an evolutionary approach*. Proceedings of the 2013 international workshop on Multi-cloud applications and federated clouds. ACM. | 2013 |
| [S112] | Ardagna, Danilo, *Modaclouds: A model-driven approach for the design and execution of applications on multiple clouds.* Proceedings of the 4th international workshop on modeling in software engineering. IEEE Press. | 2012 |

# Appendix B (Collections in the framework repository)

The catalogue of common goals that are supposed to be contributed by cloud computing technology

| Goal id | Quality goal | Explanation (from cloud service consumer perspective) | Study |
|---|---|---|---|
| G1 | Availability | Anywhere/anytime/any device (desktop, laptop, and mobile) access to resources (e.g. CPU, storage, virtual machines, and network bandwidth) which are redundant and guarantee more availability (24/7/365 and 99.99% availability) compared to run in-house infrastructure. | [S2], [S3], [S4], [S5], [S35], [S36], and [S37]. |
| G2 | Scalability | On the fly scaling up/ down resources and capability to provide varying resource demanding patterns. | |
| G3 | Security | Providing secure services protected from unauthorized access by other tenants. | |
| G4 | Performance | An excellent throughout speed and computations on cutting edge infrastructure. | |
| G5 | Customizability | Customisable and modifiable services upon requirements of consumers. | |
| G6 | Interoperability | Cloud services are integrable and incorporable with software systems as required. | |
| G7 | Portability | Systems can move from one cloud to another cloud to get better offer (e.g. performance, price, and security) with minimum disruption. | |
| G8 | Testability | Providing a scalable infrastructure to perform test and evaluation of high-computational tasks. | |
| G9 | Consistency | Guarantee of data consistency and not resulting in an error state for the system once data are processing and changing in the cloud. | |
| G10 | Reduced IT cost | Lower expense for infrastructure procuring, data storages, system updates, maintenance, and staff. | |

Probable obstacles against goals in migrating legacy systems to cloud platforms

| | | | Quality goals | | | | | | | | | | Migration type* | | | | | |
|---|---|---|---|---|---|---|---|---|---|---|---|---|---|---|---|---|---|
| # | Obstacle | Definition | Availability | Scalability | Security | Performance | Customizability | Interoperability | Portability | Testability | Consistency | Reduced IT cost | I | II | III | IV | V | Study |
| O1 | Cloud outage | A cloud service may suffer from outages for reasons such as going out of business, being the subject of regulatory action, or the outage of contact system. | * | | | | | | | | | | √ | √ | √ | √ | √ | [S2], [S44], [S58] |
| O2 | Service failure | Cloud service maybe unavailable by service consumer due to reasons such as network congestion, hardware failure, service middleware failure, or faults on various elements of the service platform. | * | | | | | | | | | | √ | √ | √ | √ | √ | [S2], [S9] |
| O3 | Service transient fault | Cloud service maybe temporarily unavailable due to network traffic load or restarting by administrators after a failure. | * | | | | | | | | | | √ | √ | √ | √ | √ | [S2], [S9] |
| O4 | Tenant interfere | Several tenants maybe in run on the same cloud and negatively affect the system data security. | | | * | | | | | | | | √ | √ | √ | √ | √ | [S59], [S60], [S88] |
| O5 | Un-customisable scalability | The scalability rules may not be flexible and merely controlled and managed by service provider. | | * | | | * | | | | | | - | √ | - | - | - | [S10], [S11] |
| O6 | Scaling latency | Cloud service may have delay in providing resource requested by service consumer due to reasons such as a server workload in the region, the rate of load acceleration, or quotas imposed by the cloud service provider. | | * | | | | | | | | | √ | √ | - | √ | √ | [S12], [S13], [S14] |
| O7 | Browser vulnerabilities | Cloud consumer who connects to cloud services by a Web browser might be attacked by malicious tenants. | | | * | | | | | | | | √ | √ | √ | √ | √ | [S15], [S16] |
| O8 | Code disruption | System codes that are executing in the cloud maybe accessed and disrupted by other tenants are in operation in the same cloud service. | | | * | | | | | | | | √ | - | - | - | √ | [S6], [S17] |
| O9 | Cloud attack | Malicious tenants can disrupt cloud service functionalities. | | | * | | | | | | | | √ | √ | √ | √ | √ | [S16], [S17], [S45], [S58] |
| O10 | Extra security cost | There might be an extra cost to address security if system components are deployed across different cloud server with complex relationships and security configuration, which demands provider-independent techniques to establish a security and configuration context. Service consumer might be responsible for locking ports, patching the operating system, running an anti-virus software and enforcement of access control policies. | | | | | | | | | | * | √ | √ | √ | √ | √ | [S6], [S26], [S28] |
| O11 | Lack of control on code execution location | Executing of a system in the cloud might not be fixed to a geographical location and rather the system may move from one physical server to another one during its lifetime. The decision on the execution location | | | * | | | | | | | | √ | - | - | - | √ | [S17], [S19] |

| | | | | | | | | | | | | | | |
|---|---|---|---|---|---|---|---|---|---|---|---|---|---|---|
| | | of the system is based on factors such as load balancing mechanism of cloud, network and server performance and availability, and even characteristics of the current consumer. | | | | | | | | | | | | |
| O12 | Lack of control on data location | Sensitive data may move to the outside the organization network or country. There is no assumption where the location of the data is. | | * | | | | | - | √ | √ | √ | √ | [S12], [S20], [S21] |
| O13 | Data remanence | The residual representation of data after finishing system execution on the cloud server may cause unwilling disclosure of private data. | | * | | | | | - | √ | √ | √ | √ | [S22] |
| O14 | Data interruption | Tenants or subcontractors of cloud providers may get access to system data and affect data confidentiality. | | * | | | | | - | √ | √ | √ | √ | [S29] |
| O15 | Session hijacking | A malicious tenant may use a valid session key to get authorised access to use system using cloud service. | | * | | | | | √ | √ | - | - | √ | [S29] |
| O16 | System source codes propriety | Cloud provider, its subcontractors, or tenant may get access to all system codes/algorithms which might be confidential. | | * | | | | | √ | √ | - | - | √ | [S17] |
| O17 | Vendor lock-in | System owner is dissatisfied with cloud service but it cannot easily and inexpensively transfer its system and data to another platform or in-house. | | | | | * | * | √ | √ | √ | √ | √ | [S50], [S51], [S52] |
| O18 | Traversal vulnerability | A malicious tenant may damage resources that are used by other tenants. | | * | | | | | √ | √ | √ | √ | √ | [S59], [S60], [S61], [S62] |
| O19 | Incompatible pluggable cloud services | At runtime, system might be plugged to a cloud service which is incompatible with the other cloud services. | | | | | * | * | √ | - | - | - | √ | [S23] |
| O20 | Incomplete APIs | Cloud service provider lacks providing a rich set of APIs. | | | | * | * | * | √ | √ | - | √ | √ | [S24] |
| O21 | Incompatible data types | Data types used in legacy and cloud service are incompatible. | | | | | * | * | √ | √ | - | √ | √ | [S12], [S38] |
| O22 | Operating system incompatibility | System components are distributed and moved among cloud servers with different operating systems which might be incompatible for managing, representing, and formatting virtual machines. | | | | | * | * | √ | - | - | - | √ | [S25], [S26], [S27] |
| O23 | Machine-image incompatibility | Virtual machines are moving between different cloud platforms but each platform has different underlying implementation for virtual machines. | | | | | * | * | √ | - | - | - | √ | [S39], [S40] |
| Q24 | Virtual machine contextualization incompatibility | Virtual machines are moving between different platforms but each platform may use different methods for customizing the context of virtual machine such as setting the operating system's username and password. | | | | | * | * | √ | √ | - | - | √ | [S39],[S40] |
| O25 | API incompatibility across multiple cloud | Cloud service may offer APIs to implement systems or virtual machines which might be incompatible with each other services. | | | | | * | * | √ | - | - | - | √ | [S25], [S26], [S27], [S30], [S40] |
| O26 | Message passing | Message passing between system and cloud services or among system components deployed on cloud servers might be unsecure and accessed by malicious tenants. Also, message size might be large affecting system performance. | | * | * | | | | √ | √ | √ | √ | √ | [S12], [S45] |
| O27 | Performance variability of cloud service | Workload variability, virtualization overheads, or resource time-sharing of cloud server may have negative effect on the system performance operating in the cloud. | | | * | | | | √ | √ | - | - | √ | [S12], [S31], [S32], [S93] |
| O28 | Geographical distance | High distance between system components that are distributed and | * | | * | | | | √ | √ | √ | √ | √ | [S12] |

| | | | | | | | | | | | | | | | | |
|---|---|---|---|---|---|---|---|---|---|---|---|---|---|---|---|---|
| | | deployed on cloud servers may cause increased latency when accessing or manipulating the data. | | | | | | | | | | | | | | | |
| O29 | Low middleware performance | A cloud service may have been built on several layers of middleware, from the guest operating system of the VM to the data-centre resource manager, which each middleware may impact on the system efficiency. | * | | * | | | | * | | | √ | √ | - | √ | √ | [S32] |
| O30 | High cancellation fees | Cloud service provider may force a consumer to a long term commitment and consumers' early exit may causes forfeit. | | | | | | | | | * | √ | √ | √ | √ | √ | [S33] |
| O31 | Inflexible pricing model | Cloud service provider may not offer a billing model based on the service usage and limit consumer to flat rates or usage thresholds. | | | | * | | | | | | √ | √ | √ | √ | √ | [S34] |
| O32 | Extra testing effort | The test of system which may be deployed on multiple cloud servers may needs testing connectivity of local components and those deployed on cloud servers along with adding a new dimension of test such as elasticity, multi-tenancy, interoperability, and elasticity. | | | | * | | | * | | | √ | √ | √ | √ | √ | [S37], [S41], [S42], [S95] |
| O33 | Learning curve | Learning a new programming style, concepts, APIs, tools, and understanding organisational impact of the cloud technology might be time consuming. | | | | | | | | | * | √ | √ | √ | √ | √ | [S43] |
| O34 | Loose of control over resources and updates | Loss of control over resource management and their update. | | | | * | | | | | | √ | √ | √ | √ | √ | [S45], [S46] |
| O35 | Bargaining power of provider | Cloud provider may get bargaining power in the future for example by raising service fee prices or refusing to invest maintenance backward compatible interface. | | | | * | | | | | | √ | √ | √ | √ | √ | [S48], [S49] |
| O36 | Proprietary APIs | Proprietary cloud APIs may impede integration of cloud services with legacy systems. | | | | * | * | * | | | | √ | √ | √ | √ | √ | [S53], [S54] |
| O37 | Licensing issue | Software is charged per instance model but cloud server creates several instances in the case of workload occurrence which might be contradictory with software licensing. | | | | | | | | | * | √ | √ | - | √ | √ | [S45], [S55] |
| O38 | Department downsizing | The maintenance team of legacy systems may become downsize as some of their responsibilities are outsourced to cloud providers. | | | | | | | | | * | √ | √ | √ | √ | √ | [S44], [S56] |
| O39 | Resistance to change | Users/staff may resist against moving to the cloud due to change in their positions and organisational structure. | | | | | | | | | * | √ | √ | √ | √ | √ | [S36], [S40] |
| O40 | Non-compliancy | Users or standard regulations don't consent to move personal/organisational data to the cloud. | | | | | | | | | * | √ | √ | √ | √ | √ | [S57] |
| O41 | Extra management effort | Maintaining a system deployed in several clouds takes extra effort such as keeping relationships with cloud providers, change of providers, and monitoring. | | | | | | | | | * | √ | √ | √ | √ | √ | [S44] |
| O42 | Backward incompatibility | System might not be easily switched between on-premise and cloud environments. | | | | | | * | | | | √ | √ | √ | √ | √ | [S77] |
| O43 | State-based dependency | System may heavily depend on contextual data, storing on server or client, such as configuration changes to operate and remain consistent from one session to another one. | * | | * | | | | | | | √ | √ | - | - | √ | [S71], [S91] |
| O44 | Incompatible APIs | Legacy system APIs and cloud's APIs are incompatible. | | | | | * | * | | | * | √ | √ | √ | √ | √ | [S12], [43] |
| O45 | Network latency | Connection speed between on premise and cloud is low due to latency in on-premise network or latency of internal cloud network. | * | | * | | | | | | | √ | √ | √ | √ | √ | [S12] |
| O46 | Browser latency | The browser in the on-premise environment is working slowly. | * | | * | | | | | | | √ | √ | √ | √ | √ | [S12] |
| O47 | Service latency | Latency in performing cloud service due to obstacles O7, O28, O46, and O45. | * | | * | | | | | | | √ | √ | √ | √ | √ | [S12] |
| O48 | Incompatibility of legacy system and cloud service | Incompatibility between legacy system and cloud services due to obstacles O21, O22, O23, Q24, and O25. | | | | | * | * | | | * | √ | √ | √ | √ | √ | [S12], [S43] |
| O49 | Incompatibility of legacy system data storage and cloud | Incompatibility between legacy data storage and cloud database solution due to O21 and O50. | | | | | * | * | | | * | - | - | - | √ | - | [S12], [S43] |
| O50 | Incompatible data operations | Stored procedures, views, and functions providing by cloud data store might not be compatible (either syntactically or semantically) with those defined in legacy system. | | | | | * | * | | | * | - | - | - | √ | - | [S12], [S43] |
| O51 | Tight dependencies | Tight dependencies among legacy system components or dependency to underlying technologies, operating systems, programming language, or other legacy systems may obstruct individual scalability and portability of system components across multiple clouds and on premise. | * | | | | | * | * | | | √ | √ | √ | √ | √ | [S67], [S98], [S108] |
| O52 | Inconsistency of system components | Cloud data storage services may offer weaker consistency properties in the sense that it will be taken long time to have consistent data across all servers. | | | | | | | | * | | - | - | √ | √ | √ | [S8], [S18] |
| O53 | Identity theft | An attacker may get a valid user's identity and access resources of legacy systems. | | * | | | | | | | | √ | √ | √ | √ | √ | [S104] |
| O54 | Variable price of cloud resources | The price of using cloud resources may vary depending on cloud workload across the time, particularly in a pick period. Such price variation may not be suitable for legacy systems with heavy processing tasks. | | | | | | | | | * | √ | √ | - | - | √ | [S112] |
| O55 | High cost of support (Specific to AWS) | AWS is a general provider which expects its services to be used and managed by its uses independently. If a problem occurs, AWS has an expensive technical support. | | | | | | | | | * | √ | - | - | - | - | [S102] |
| O56 | Unreliable IT support (Specific to AWS) | The quality of IT support by AWS might be a risk. | | | | | | | | | * | √ | - | √ | - | √ | [S100] |
| O57 | Varying support fee (Specific to AWS) | AWS support fees vary on a sliding scale tied to monthly in a way that support costs may grow quickly if system performs heavy tasks. | | | | | | | | | * | √ | - | - | - | - | [S101] |
| O58 | Vulnerable security (Specific to AWS) | Amazon simple storage service (S3) may be accessible via SSL (secure socket layer) encrypted end points, implying that it is the user's responsibility to encrypt data before storing into S3. | | * | | | | | | | | - | - | √ | - | - | [S103] |
| O59 | Injection attack (Specific to AWS) | An attacker may hijack user accounts by creating, modifying, and deleting virtual machine images, and changing administrative passwords to control interfaces used to manage cloud computing resources (e.g. S3 or EC2). | | * | | | | | | | | √ | - | √ | - | - | [S104] |
| O60 | Inflexible cost model (Specific to Azure) | Azure computes the cost of recourses that were used per minute with rounding up service usage to the nearest minute. In other words, if a user allocated a resource for one hour and a half, then payment is computed for the exact period of time whilst a provider like Amazon round up service consumption to nearest hour. | | | | | | | | | | * | √ | √ | √ | √ | √ | [S99] |
| O61 | Inflexible configuration (Specific to Azure) | The provider may not provide high flexible hardware configurability for each virtual machine instance compared to Amazon offering high flexibility in virtual machine configuration. | | | | | * | | | | | | √ | - | - | - | - | [S104] |
| O62 | Operating system incompatibility (Specific to Azure) | Microsoft Azure mainly supports Windows-based servers. Porting legacy systems from other platforms (e.g. Linux) to Azure might require modifying the source code to be compatible with Windows APIs and them able to execute on Azure. | | | | | * | | | | | | √ | | | | | [S104], [S106] |
| O63 | Limited geographical zone (Specific to Google) | Google may not provide an extensive coverage of data centres to deploy legacy systems. | * | | | | | | | | | | √ | | | | | [S107] |
| O64 | Inflexible cost model (Specific to Rackspace) | Rackspace may offer limited pricing options and month-to-month subscriptions. | | | | | | | | | | * | √ | - | - | - | - | [S102] |
| O65 | Heterogeneous production environments | The complexity and differences between production environments (e.g. cloud platform, third-party clouds, and legacy systems) related to deployments and configurations can hinder the efficiency of test. | | | | | | | | * | | | √ | √ | √ | √ | √ | [S108], [S109] |
| O66 | Costly virtual machine | Virtual machine and its underlying infrastructure might be costly in terms of need for large disk storage, isolated binary and library files, memory management, and full gust operating system image. | | | | | | | | | | * | √ | √ | - | - | √ | [S110] |
| O67 | Incompatible execution environments of system | A legacy system which is encapsulated in a virtual machine may not be interoperable and portable across multiple cloud platforms. | | | | | | * | | | | | √ | √ | - | - | √ | [S110] |

* For the migration types see the migration criteria in Section 6.

Catalogue of resolution tactics for handling obstacle

| # | Resolution tactic | Definition | Relation to obstacle | Source | Category |
|---|---|---|---|---|---|
| T1 | Substitute goal | Identify an alternative goal which is still contributable by the chosen migration type or cloud services in a way that the obstructed goal and obstacle will not occur. | Applicable to resolve all obstacles | KAOS framework | Goal/Service/Migration type Substitution |
| T2 | Substitute cloud migration type | Choose an alternative cloud adoption type which satisfies the obstructed goal is adopted in a way that the obstacle will no longer occur. The tactics has root in the fact that different cloud adoption types, besides their specific contributions to quality goals, might have common contributions towards migration goals. | Applicable to resolve all obstacles | Adopted from KAOS framework | Goal/Service/Migration type Substitution, Obstacle reduction |
| T3 | Substitute cloud service | Resolve the obstacle by selecting/changing the cloud service/provider in a way that new the cloud service can contribute to quality goals. Define a set of suitability criteria that characterise desirable features of cloud providers. The criteria include provider profile (e.g. pricing model, constraints, offered QoS, electricity costs, power, and cooling costs), organisation migration characteristics (migration goals, available budget), and system requirements. Based on the criteria, identify and | Applicable to resolve all obstacles | Adopted from KAOS framework and [S65], [S79] | Goal/Service/Migration type Substitution, Obstacle reduction |
| | | select suitable cloud providers. | | | |
| T4 | Analyse migration feasibility | Perform a feasibility analysis to evaluate the benefits and the consequences of moving legacies to the cloud and its impact on organisation structure, staff's roles, and legacies. | O38, O39 | [S73], [S74] | Obstacle prevention |
| T5 | Refactor legacy source code | Adapt the source code for being compatible and able to interact with the selected cloud platform programming language and APIs. | O19, O20, O21, O22, O23, Q24, O25 | [S65], [S66], [S67] | Obstacle prevention |
| T6 | Develop adaptor/wrapper | Add adaptors for resolving mismatches, occurring at runtime system execution, between legacy system components and cloud services. | O19, O20, O21, O22, O23, Q24, O25, O36, O50 | [S75], [S76] | Obstacle prevention |
| T7 | Decouple system components | Decouple the legacy system components from each other. Use mediator and synchronisation mechanisms to manage interaction between the loosely coupled components in the cloud environment. | O51 | [S12], [S77], [S78] | Obstacle prevention |
| T8 | Encrypt/decrypt message passing | Add support for the runtime encryption/decryption of message transition between components in on-premises network and cloud environment. | O26 | [S12], [S75], [S79] | Obstacle prevention |
| T9 | Obfuscate code | Protect unauthorised access to code blocks of components by other tenants that are running on the same cloud service. Use encryption mechanisms in the sense that no other tenants will be able to access, read, or alter the code blocks with the components when running in the cloud. | O8, O16 | [S6] | Obstacle prevention |
| T10 | Isolate tenant | Enable multi-tenancy in the system. Based on multi-tenancy requirement (i) define tenant-based identification and hierarchical access control for tenants and (ii) separate tenant data using authorization and authentication mechanisms. | O4 | [S80], [S81], [S96], [S97] | Obstacle prevention |
| T11 | Tune message granularity | Define suitable granularity for messages, that are passing between system components hosted on local network and the cloud, based on the degree of functionality that is offered to the service consumer and consumer's infrastructure capability to process the messages. A proper message granularity can be identified or predicted based on pieces of data actually used by system or using heuristic functions to understand the number of interaction between system components over the cloud network. | O26 | [S12], [S82] | Obstacle prevention |
| T12 | Adapt data | Convert legacy data types to the data type of target cloud database solution. Also, add an extension component to the legacy system which includes a set of commands to be performed by the system or cloud. The emulator supports missed database functionalities of cloud database solution provider. | O50, O21 | [S12], [S38], [S71], [S83], [S84] | Obstacle prevention |
| T13 | Involve staff with cloud adoption process | Involve staff and stakeholders actively in the cloud adoption process and give them insight of benefits of the cloud and organisational change. | O38, O39 | [S46] | Obstacle reduction |
| T14 | Define an authorization | Add a component determining if a tenant has privilege to perform a given action over the database. | O4 | [S69] | Obstacle prevention |
| T15 | Encrypt data | Use data encryption mechanisms prior outsourcing or hosting system data to the cloud. | O14, O13, O4 | [S12], [S79], [S85] | Obstacle prevention |
| T16 | Filter unauthorised requests | Add support to filter unauthorized data access received from users at the edge of premise or cloud network as early as possible to avoid unauthorized network traffic. | O14, O4 | [S58] | Obstacle prevention |
| T17 | Adjust security policies | Add support for runtime security assessment of received queries for run on data. | O14, O4 | [S58] | Obstacle prevention |
| T18 | Replicate system components | Partition, replicate, and distribute system components and data (replicas) on multiple cloud servers. | O3, O6, O27, O45, O28 | [S58], [S78], [S86] | Obstacle reduction |
| T19 | Backup periodically | Implement a procedure to periodically perform data backup. | O4, O14, O15, O17 | [S71], [S72] | Obstacle prevention |
| T20 | Detect and filter intrusions | Filter unauthorised packets and malformed data traversed between system components in local network and the cloud environment. | O4, O8, O9 | [S58], [S70] | Obstacle prevention |
| T21 | Update patches | Perform regular patch update across system components in the cloud. | O7, O8, O9, O46 | [S74], [S87] | Obstacle reduction |
| T22 | Isolate tenant | Protect tenants' data from to be accessed by other tenants. Each tenant should be authorised and able to access to its own data. | T7 | [S88], [S89], [S90] | Obstacle prevention |
| T23 | Define retry policies | Define retry policies and implement them in the system for the operation to succeed. | O3 | [S66], [S75] | Goal restoration |
| T24 | Refine network topology | Define a proper network topology with a consideration of server proximity and system components, proper provider equipment, the location of the data centres, router hops, and infrastructure bandwidth. | O27, O28, O47, O45 | [S65], [S66], [S74], [S77], [S78] | Obstacle reduction |
| T25 | Examine cloud service behaviour | Use benchmarking tools to investigate performance of the cloud under investigation before decision making. | O27, O11, O12, O17 | [S32], [S65] | Obstacle prevention |
| T26 | Acquire more cloud resources | Rent more VMs or higher spec ones to deal with slow CPU clock rates, use physical disk shipping to reduce effects of network latency/transfer rates. | O27 | [S2], [S92] | Obstacle reduction |
| T27 | Use multiple cloud servers | Deploy and replicate system components in several clouds. | O27 | [S45] | Obstacle reduction |
| T28 | Add intermediation | Implement an intermediate layer (mediator components) between legacy system and cloud services that decouple legacy systems from cloud specific APIs. This helps to create intermediate APIs and get indirect service from the cloud. | O6, O29, O47 | [S63], [S64] | Obstacle prevention |
| T29 | Make system stateless | Provide a support in the system to the handle safety and traceability of tenant's session when various system instances are hosted in the cloud. | O43 | [S78], [S91] | Obstacle prevention |
| T30 | Prioritize tests | Perform test cases on the basis of their importance and criticality. | O32 | [S95] | Obstacle prevention |
| T31 | Resolve licensing issue | There are alternative sub-tactics: (i) negotiate with system owner to make a suitable licensing model which satisfies all parties, (ii) extend legacy system with a new component (e.g. VPN tunnel) in a way that cloud services can be indirectly offered to them, and (iii) enable a license tracking mechanism through monitoring connections between the software system and cloud resources. | O37 | [S72], [S74] | Obstacle prevention |
| T32 | Define weak inconsistency | Implement an eventual consistency or similar weak consistency model for data. | O52 | [S8] | Obstacle reduction |
| T33 | Check compliance | Check if cloud adoption is compliance with the auditors and cloud providers. | O40 | [S45], [S57] | Obstacle prevention |
| T34 | Clarify roles | Clarify roles and responsibilities relevant to cloud adoption. | O38, O39 | [S40], [S45] | Obstacle reduction |
| T35 | Aware top-level management | Make management aware of the extra effort that might be required for cloud adoption in the organisation. | O31, O33, O35, O38, O39, O41 | [S94] | Obstacle reduction |
| T36 | Degrade goal | Resolve an obstacle by degrading goal definition and refining its assumption for required levels of satisfaction so that the refined goal makes more freedom for violation. | Applicable to resolve all obstacles | KAOS framework | Goal weakening |
| T37 | Restore goal | Add a new goal for restoring the satisfaction of the obstructed goal when violated. | Applicable to resolve | KAOS framework | Goal restoration |
| | | | all obstacles | | |
| T38 | Mitigate goal | Add a new goal for mitigating the consequences of an obstacle if it occurs. | Applicable to resolve all obstacles | KAOS framework | Goal mitigation |
| T39 | Fix inconsistencies | Perform manual or semi-automate steps to resolve inconsistencies which have occurred after data operations. | O52 | [S8] | Goal mitigation |
| T40 | Define compensation | Specify penalties (e.g. financial or getting more quote) to be paid by cloud provider in the case of a disruption. | O1, O2 | [S2], [S3], [S4] | Goal mitigation |
| T41 | Do nothing | Leave obstacle unresolved. | Applicable to resolve all obstacles | KAOS framework | Do nothing |
| T42 | Use rigorous authentication | Use strong passwords and authentication mechanism when running system in cloud environment. | O53 | [S104] | Obstacle prevention |
| T43 | Keep virtualization at the system level | Create virtualization and isolation boundary at the legacy system level rather than at the server level through container concept. Such a container (i) handles resource allocation meaning that in the case of excessive resource consumption by a system operating in the cloud, only individual container is affected and whole virtual machine is left unaffected and (ii) reduces incompatibility problems between systems across multiple platforms. | O24, O66, O67 | [S110] | Obstacle prevention |
| T44 | Use dedicated virtual machine | Run the system on dedicated virtual machine in the sense that the virtual machine is entirely performed on separate resources such physical servers, network, switch, bandwidth, disk, CPU, memory to satisfy expected goals, i.e. quality of service. All resources are physically dedicated to the virtual machine. | O4, O6, O8, O9, O13, O14, O18, O27, O29, O53, O58, O59 | [S111] | Obstacle reduction |
| T45 | Define bidding strategy | Identify heavy processing tasks of the system (e.g. image, video, conversion and rendering) and define a bidding strategy for spot instance to lessen the cost of using cloud resources. | O54 | [S68] | Obstacle prevention |

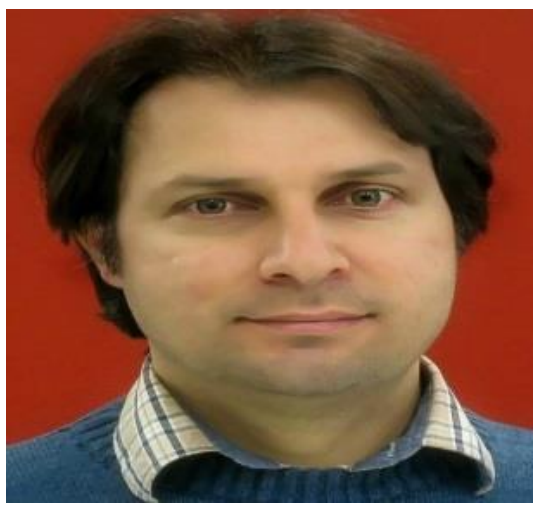

Mahdi received a PhD degree in Information Systems from the University of New South Wales, Sydney. He also holds a master degree in software engineering. He is currently a research associate at University of Technology Sydney. Mahdi's vision in research is to develop IT-based solutions for real problems or to help organizations in adopting IT innovations in a systematic way. His research interests lie in the areas of cloud computing, big data, conceptual modelling, and method engineering. Mahdi has hands-on experience in design and development IT solutions in different industry domains including accounting, insurance, defense, and publishing.

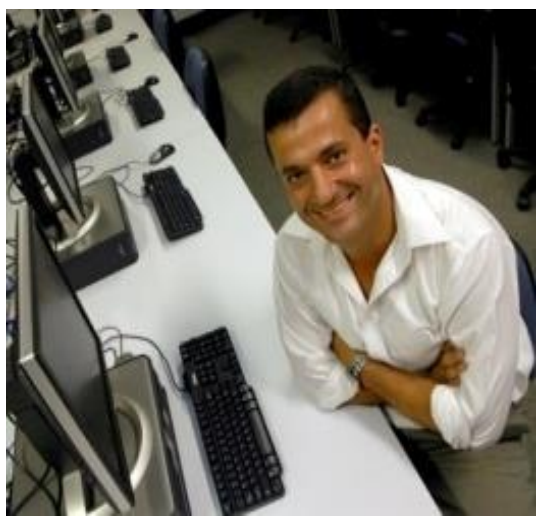

Professor Ghassan Beydoun received a degree in computer science and a PhD degree in knowledge systems from the University of New South Wales, Sydney. He is currently a Professor of Information Systems at the University of Technology Sydney. He has authored more than 100 papers international journals and conferences. He is currently working on the metamodels for on project sponsored by Australian Research Council and Australian companies to investigate the endowing methodologies for distributed intelligent systems and supply chains with intelligence. His other research interests include multi agent systems applications, ontologies and their applications, and knowledge acquisition.